\documentclass[final,5p,times,twocolumn]{elsarticle}

\usepackage{graphicx}
\usepackage{amssymb}
\usepackage{float}
\usepackage{placeins}
\usepackage{amsmath}
\usepackage[table]{xcolor}
\usepackage{array}
\usepackage{multirow}
\usepackage{booktabs}
\usepackage[group-separator={,}]{siunitx}

\newcommand\MyBox[2]{
  \fbox{\lower0.75cm
    \vbox to 1.7cm{\vfil
      \hbox to 1.7cm{\hfil\parbox{1.4cm}{#1\\#2}\hfil}
      \vfil}%
  }%
}

\usepackage{lineno}

\journal{Journal of Finance and Data Science}
\begin{document}

\begin{frontmatter}


\title{Validating Weak-form Market Efficiency in United States Stock Markets with Trend Deterministic Price Data and Machine Learning}



\author{Samuel Showalter, Jeffrey Gropp}

\address{Department of Economics, DePauw University, Greencastle, IN 46135, USA}

\begin{abstract}
The Efficient Market Hypothesis has been a staple of economics research for decades. In particular, weak-form market efficiency -- the notion that past prices cannot predict future performance -- is strongly supported by econometric evidence. In contrast, machine learning algorithms implemented to predict stock price have been touted, to varying degrees, as successful. Moreover, some data scientists boast the ability to garner above-market returns using price data alone. This study endeavors to connect existing econometric research on weak-form efficient markets with data science innovations in algorithmic trading. First, a traditional exploration of stationarity in stock index prices over the past decade is conducted with Augmented Dickey-Fuller and Variance Ratio tests. Then, an algorithmic trading platform is implemented with the use of five machine learning algorithms. Econometric findings identify potential stationarity, hinting technical evaluation may be possible, though algorithmic trading results find little predictive power in any machine learning model, even when using trend-specific metrics. Accounting for  transaction costs and risk, no system achieved above-market returns consistently. Our findings reinforce the validity of weak-form market efficiency.
\end{abstract}

\begin{keyword}
Efficient Market Hypothesis \sep Support Vector Machine \sep Random Forest
\sep Bayesian Theory\sep K-Nearest Neighbors \sep Logistic Regression \sep Moving Average \sep Dickey-Fuller Test \sep Stationarity \sep Unit Root \sep Variance Ratio


\end{keyword}

\end{frontmatter}



\section{Introduction}
\label{S:1}

Stock market efficiency has been proposed, tested, and debated for nearly a century. Three versions of the Efficient Market Hypothesis have arisen from an immense body of economic and econometric research. For a market to be deemed efficient, prices must reflect currently available information \cite{Fama1970}. In its weakest form, available information is defined to be historical prices. More stringent theories like the semi-strong EMH include all publicly available information, while the strongest form adds to that information that is not publicly known \cite{Fama1970}. 

While efficient market research gained wide-spread attention in the 1960s and 1970s, these theories originated nearly a century prior. In his book \textit{Theory of Speculation}, Bachelier \cite{bachelier1900theorie} first proposed the notion of efficient markets as the concept of ``fair game" economics. The expected return of the speculator is zero. \cite{bachelier1900theorie} In fact, his characterization of stock price movements as a drunkard's errant walk coined the term ``Random Walk", used to name the Random Walk Theory \cite{DeBondt1989}.  

Today, many econometric and statistical techniques are employed to validate efficient market behavior. In particular, unit root tests discern whether or not time series data is stationary \cite{dickey1979distribution}. Stationarity characterizes a time-series whose mean, variance, autocorrelation, and other traits remain constant over time, implying the presence of trending and potential exploitation opportunities. Conversely, a unit root represents a time series that demonstrates systematic randomness, and can be one cause for not seeing stationarity in data \cite{phillips1988testing}. Methods for testing for a unit root include, but are not limited to, \cite{cheung1995lag,phillips1988testing,dickey1979distribution} Dickey-Fuller, Phillips-Perron, and Kwiatkowski-–Phillips–-Schmidt–-Shin (KPSS) tests. At the same time, alternative tests of the weak-form EMH include the Variance Ratio, Serial Correlation, and Runs tests  \cite{borges2010efficient}.

Evidence supporting efficient market hypotheses from the 1960s went relatively undisputed for decades. Though the EMH (henceforth referencing weak-form EMH)
has been tested and supported by many economists, Eugene Fama is considered the pioneer of efficient market research. In his 1965 dissertation \cite{Fama1965}, Fama extensively examines the characteristics of stock price movement, ultimately publishing what he and many defer to as ``strong and voluminous evidence in favor of the random-walk hypothesis." His work identifying the nature of speculation and price movement continues well into the $21^{st}$ century, often in collaboration with Kenneth French \cite{FamaFrench2004}. 

Researchers aligned with efficient market theories experiment with different methods, but often converge to the same conclusion: one cannot generate above-market returns through speculation or ``timing the market" when attributes such as risk, transaction costs, and expenses are considered \cite{malkiel1991random}. Non-stationarity in stock data implies -- but does not prove -- an inability to predict future price movement and generate above-market returns through speculation. One example of a historical trending anomaly proven to not be consistently predictive is the January effect, where stock prices consistently rise in January every year \cite{malkiel1991random}. The Small Firm Effect similarly asserts that smaller firms outperform larger companies, though little evidence verifies its validity. Other, more far-fetched occurrences include the Hemline Effect, where the popular hemline length of women's clothing indicates the returns of the market \cite{malkiel1991random}. 

Dominance of EMH theory peaked in the 1970s \cite{shiller2003efficient}, as research opposing the Efficient Market Hypothesis gained traction around the 1980s. At this time,  evidence countering market efficiency grew, but through different lines of reasoning. For example, Modigliani et. al. (1979) regard asset prices as chronically undervalued \cite{modigliani1979inflation} due to inflation, while others attribute market inefficiency to irrationality on behalf of the economic participant \cite{DeBondt1989}. Groundbreaking research on the psychology of choice finds people inherently and consistently act irrationally when influenced by certain environments \cite{tversky1981framing}. Even so, some maintain that consistently achieving above-market returns is still unlikely when returns are adjusted for risk and fees, despite evidence of market trending \cite{banz1981relationship} . In this regard, stock market inefficiency and speculative arbitrage are separate matters entirely. Establishing markets are inefficient does not inherently imply that above average returns can be gleaned through arbitrage.

Today, yet another field is influencing the validity of the weak-form EMH. Innovations in technology have given rise to artificially intelligent systems. In particular, the influence of machine learning -- a discipline dedicated to algorithms that "learn" to classify or pattern match -- is altering the nature of business and academia alike \cite{witten2016data}. One field particularly affected is financial engineering. The use of algorithmic trading platforms seeking arbitrage opportunities has proliferated, as have publications detailing methods of attracting "abnormal returns" \cite{Dash2016,Kuo2001}. Some analyses assert their trading algorithms are able to outperform the market outright by training models on price data alone \cite{Patel2015}.

\section{Contribution}
\label{S:2}

This paper seeks to merge independent silos of research in economics and machine learning, ultimately to garner a holistic perspective on the validity of the weak-form Efficient Market Hypothesis. While debate on efficient markets persists in economics, few contest the weak-form EMH due to its conservative nature \cite{Fama1998}. In general, economist sentiment in academia affirms the weak-form Efficient Market Hypothesis, and many make such an assumption when conducting research unrelated to market efficiency \cite{Summers1986}. 

By contrast, machine learning research has found little consensus in its findings, partially due to the wide variety of implementations used to predict stock market movement. Nevertheless, there is a consistent trend of deriving above average returns from publicly available data \cite{schumaker2009textual,enke2005use,Dash2016,Patel2015}. The majority of implementations make use of information that may not be considered by the weak from EMH. Even so, studies completed by \cite{Dash2016} and \cite{Patel2015} purport to have built a system that can beat the market on price data alone. 

Perhaps machine learning solutions are correct. The value of AI is embedded in its ability to perform as well or better than humans at pattern detection \cite{nasrabadi2007pattern}. Moreover, a growing population of economists are not finding financial markets around the world to be weak-form efficient \cite{worthington2003weak,Nisar2012,buguk2003testing,poshakwale1996evidence}. Granted, some of these markets are considered to be "emerging" and not fully developed, as corroborated by research on countries in socio-political turmoil.

Thus, to control for as many extraneous factors as possible and also test the weak-form EMH, we first implement two econometric tests for market efficiency. Augmented Dickey-Fuller (ADF) and Lo-Mackinlay \cite{lo1988stock} variance ratio tests are commonly used to support and refute the Random Walk Hypothesis, respectively. While each does not explicitly indicate whether above-market returns are possible through speculation or arbitrage, they provide a framework from which further exploration can be conducted. There may be opportunities to exploit market trends to gain a profit if econometric studies indicate as much.

To logically transition between stationarity and market returns, an algorithmic trading platform is built and implemented. Daily stock index price and volume data from developed American financial markets -- IXIC, GSPC, RUT, and DJI
-- is collected. Then, 10 technical indicators identical to those collected by \cite{Patel2015} are derived in the same manner. Trend-specific metrics that convert ratios into their buy sell intuition are also generated.

Stock data is fed into a machine learning platform that predicts the following day's stock price momentum (down or up, where up includes breaking even
) 
using the technical ratios from the previous day. After each prediction is made, the de-facto outcome of the market is harnessed to re-train the system such that it constantly has current information. Prediction results feed a trading platform that tracks performance (\%-return, Alpha, Beta, Volatility, Sharpe, Sortino) using \$100,000 of principal to trade a specific security. Risk free and buy-and-hold returns are used as benchmarks.

Stationarity, prediction, and trading tests together provide a general perspective of  weak-form market efficiency's validity. While this is but one approach to garnering above-market returns, it provides strong evidence for characterizing the weak-form EMH. Independent verification is also conducted with 100 randomly selected stocks traded as a part of the S\&P 500.

\section{Stock and Stock Index Data Sources}
\label{S:3}

The algorithmic trading platform only uses adjusted close prices and trades according to the black demarcations in Figure 1. The blue timeline represents market open and close on a given day, and the trading platform's training data depends on the adjusted close the previous day. 
\FloatBarrier
\begin{figure}[!ht]
\label{fig:TradeDay}
\centering
\includegraphics[width = 1\linewidth]{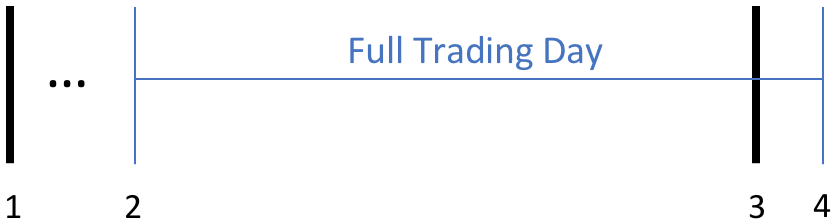}
\caption{Visualization of regular trading day compared to trade interval for platform. 2 and 4 represent market open and market close. 1 and 3 represent the moment milliseconds before market close over a two day consecutive window.}
\end{figure}
\FloatBarrier

However,  we assume the algorithm can not trade after hours. The algorithm generates a prediction and trades using the price found  during the final milliseconds before the trading day ends.
These trades stem from predictions on the market activity the following day. A maximum of one trade occurs daily, and mistakes do not alter trading decisions.

Stock index data is acquired over Yahoo Finance, while the stock price data for individual securities is procured through the Quandl WIKI API. These data sources are reputable, but prices were programmatically checked for blatant errors and missing data before any experimentation was conducted. Both of these data sources define adjusted close as the closing stock price adjusted for stock splits, reverse splits, and dividends.

\section{Quantitatively Characterizing Efficient Markets}
\label{S:4}

\subsection{Econometric Examination of Stochastic Price Trends}

Beating the market is a multi-faceted concept. Not only does it require that price movements exhibit predictability, but that variations must have the potential to be systematically exploited \cite{FamaFrench2004}. Most commonly, value investors believe that markets under- and over-value stocks based on public sentiment. Frequent exploitation of market efficiency inherently and paradoxically increases efficiency,  asserted by \cite{brogaard2014high} with high frequency trading. Therefore, markets will only maintain their inefficiency if exploited by a select few.

To simplify matters, economists often objectively test for randomness in stock data and use their findings as a proxy for both market inefficiency and the potential to generate above-market returns. Unit root tests have evolved into many different forms, partially because "most aggregate economic time series contain a unit root" \cite{kwiatkowski1992testing}. That is, most time-series in raw form appear non-stationary. However, some series can be converted into stationary systems through de-trending, differencing, and other methods \cite{breitung2001local}. Data coerced to stationarity in this manner is considered difference-stationary. If a unit root test implies the presence of stationarity, then there may exist securities with non-random price movements that can be exploited.

Variance Ratio tests rely on the assumption that the ratio of variance for two different time frames should be approximately proportional to the ratio of those lengths of time. The expansion of variance over time in non-stationary data is considered a linear process \cite{charles2009variance}. Additionally, Runs tests \cite{wald1940test} examine the length of "runs", where stock price repeatedly moves in the same direction
. In particular, it discerns whether or not price movements are mean reverting, one of the most-commonly proposed hypotheses of Random Walk opponents \cite{DeBondt1989}.

\subsection{Automated Trading and Binary Classification}

Algorithmically, price movements are considered to have either up or down momentum. Two potential outcomes predicate two possible predictions, and converts stock price prediction into a binary classification problem. As displayed in Table 1, four potential outcomes arise from a prediction-result pair.

\begin{table}[h]
\begin{tabular}{c >{}r @{\hspace{0.7em}}c @{\hspace{0.4em}}c @{\hspace{0.7em}}l}
  \noindent
  \renewcommand\arraystretch{1.5}
  \setlength\tabcolsep{0pt}
  \multirow{11}{*}{\parbox{0.7cm}{\bfseries\raggedright Prediction}} & 
    & \multicolumn{2}{c}{\bfseries Actual Price $\Delta$} & \\
  & Price Up & \MyBox{True}{Positive (TP)} & \MyBox{Price Up}{Positive (FP)} \\[2.4em]
  & {Price Down} & \MyBox{False}{Negative (FN)} & \MyBox{True}{Negative (TN)} \\
 && Price Up & Price Down &
\end{tabular}
\caption{Confusion matrix for daily stock price momentum prediction.}
\end{table}

Many assessment metrics are derived from Table 1. Listed in equations \ref{eq:TPRFNR}, \ref{eq:TNRFPR}, \ref{eq:PPVFDR}, and \ref{eq:NPVFOR} are the eight basic indicators derived from confusion matrices: true and false positive and negative rates ($TPR$, $FNR$, $TNR$, $FPR$), positive and negative predictive value ($PPV$, $NPV$), false discovery rate ($FDR$), and false omission rate ($FOR$). Visually, all equations in the left column focus on successful classification, while the right column focuses on failure rates.

\begin{equation}\label{eq:TPRFNR}
{TPR = \frac{TP}{TP + FP} \ \ \ \ \ \ \ \ \ \ FNR = \frac{FN}{FN + TN} }
\end{equation}

\begin{equation}\label{eq:TNRFPR}
{TNR = \frac{TN}{TN + FN} \ \ \ \ \ \ \ \ \ \ FPR = \frac{FP}{FP + TP} }
\end{equation}

\begin{equation}\label{eq:PPVFDR}
{PPV = \frac{TP}{TP + FN} \ \ \ \ \ \ \ \ \ \ FDR = \frac{FN}{TP + FN} }
\end{equation}

\begin{equation}\label{eq:NPVFOR}
{NPV = \frac{TN}{TN + FP} \ \ \ \ \ \ \ \ \ \ FOR = \frac{FP}{TN + FP} }
\end{equation}

\bigbreak
Conceptually, TPR and TNR determine how many price movements were appropriately classified. Both of these metrics are direct evaluations of model performance by class, and ask the question: \textit{how many price movements were classified correctly?} Conversely, PPV and NPV can be described as a $prediction \ centric$ evaluation of the confusion matrix, conditionally examining results given a specific prediction. That is, \textit{for all days predicted to have price increase, how many actually did?}  

Many higher-order measurements combine the ratios above into a holistic indicator of performance. Two particularly important for metrics are Accuracy and the $F_1$ Score \cite{Patel2015}. Accuracy is a common form of evaluation that highlights the number of days that were classified correctly. By contrast, the $F_1$ Score places excess emphasis on upward price movement.

\begin{equation}\label{eq:Acc}
{Accuracy = \frac{TP + TN}{TP + TN + FP + FN}}
\end{equation}

\begin{equation}\label{eq:F1Score}
{F_1 \ Score = \frac{2TP}{2TP + FP + FN}}
\end{equation}

Both measurements characterize the performance of different price trends, but tell little of overall trading efficacy. Therefore, portfolios are tracked with a variety of features, including the number of trades a system makes, portfolio returns, Alpha, Beta, Volatility, Sharpe, and Sortino. Defined and contextualized in \textit{Research Methods}, portfolio evaluation allows investors to understand the basis of returns as a function of different factors like risk ($\sigma$) and correlation with the market benchmark ($\beta$). Modern Portfolio Theory (MPT) \cite{Markowitz1952} involves gaining the highest return for a given level of risk. Optimally, an algorithmic trading system would abide by these tenets and exist on an efficient frontier.

\section{Efficient Market Research: Findings and Challenges}
\label{S:5}

Definitive conclusions on the weak-form Efficient Market hypothesis are difficult to ascertain. Evidence that stock prices follow a Random Walk does not support the notion that prices reflect all available information. Furthermore, asserting that stocks generally reflect all information is distinct from \textit{always} reflecting available information immediately. Modern technology \cite{brogaard2014high} allows for arbitrage to execute in microseconds, necessitating instantaneous price corrections.

When Fama first proposed the efficient market hypothesis \cite{Fama1965}, such technology did not exist. Over time, his continued research has transitioned from confident affirmation of EMH theories \cite{Fama1970} to addressing skepticisms, and finally conceding that valid studies have "uncover[ed] empirical  regularities" in price data \cite{fama1991efficient}  that provide a compelling argument for the contrapositive. Strong arguments maintain that financial markets are at least temporarily inefficient.

Leading critics of efficient market theories in academia are Lo and Mackinlay \cite{LoMacNonRand} and Richard Thaler \cite{DeBondt1989}, and the central notion of their research  is that economic agents participating in financial markets do not universally act rationally \cite{DeBondt1989}. Be it the Tulip Mania craze of the 1600s \cite{malkiel1991random} or the 2008 financial crisis, there are many examples of what appears to be irrational speculation en-masse
. Similarly, some assert empirical evidence of randomness in stock price movements is also evidence of proper asset valuation. Opponents like \cite{LoMacNonRand} are quick to deem this non-sequiturial, fallacious logic.

Complicating matters further, making money off of stock market trends
is often considered a separate debate from simply determining that stock prices do not follow a random walk. Many machine learning studies tout successful attempts to gain abnormal returns, albeit many do not include factors such as risk, transaction costs, and the like. 

Additionally, studies by \cite{Dash2016}, \cite{Kuo2001}, and others make use of sentiment data, company fundamentals, or other information that may not be considered when testing the weak-form efficient market hypothesis. However, \cite{Patel2015} and \cite{Kara2011} both train their models with technical ratios alone. In Indian markets, \cite{Patel2015} achieve accuracy in excess of 85\% for four securities as well as more than double the returns of their benchmarks. However, the limited nature of  four results invites random anomalies or security picking by the authors. No information is provided on how these four securities were chosen. Regardless, our analysis will provide evidence that refutes these findings as they pertain to U.S. markets.

%

\section{Research Methods}
\label{S:7}

\subsection{Econometric Analysis of Random Walk Theory}

To garner a general understanding of market efficiency, baseline tests were run on 100 randomly
selected S\&P 500 securities from Jan 1. 2008 to Jan. 1, 2018.
More specifically, both tests are run on the daily log returns
. Augmented Dickey Fuller (ADF) tests, one of the commonly cited tests of efficient market proponents, and Variance Ratio tests were both conducted. Lo and Mackinlay are staunch opponents of Random Walk theory \cite{LoMacNonRand} and created a specific implementation of the Variance Ratio test to verify their hypotheses \cite{lo1988stock}. Taken together, these independent analyses will both frame the ongoing random walk debate as well as the algorithmic experiments to come.

\bigbreak
\textbf{Augmented Dickey-Fuller (ADF) Tests:} ADF tests are a form of unit root analysis that incorporates lags to account for serial correlation \cite{cheung1995lag}. Over a long time period, data may appear to follow patterns when, in actuality, these movements are the product of randomness or drift
\cite{Leybourne1998}. ADF is statistically defined as examining the null hypothesis of ``an (autoregressive integrated moving average) ARIMA(p, 1,0) process against the stationary ARIMA(p + 1, 0, 0) alternative" \cite{cheung1995lag}. If $x_t$ is a time series, then ADF regression takes the form

\begin{equation}\label{eq:ADFtest}
{\Delta x_t = \mu + \gamma t + \alpha x_{t-1} + \sum_{j=1}^{k-1} {\beta_j \Delta x_{t-j} + \mu_t}}
\end{equation}

where  ``$\Delta$ is the difference operator and $\mu_t$, is a white-noise innovation" \cite{cheung1995lag}. ADF tests in this study were automated using the $urca$ package of R. The number of lags included in the analysis was automatically determined using the Akaike Information Criteria (AIC), and the maximum number of lags $p_{max}$ considered was based on the paradigm of \cite{schwert1989tests} outlined in equation 8.

\begin{equation}\label{eq:SchwertLags}
{p_{max} = 12\left(\frac{T}{100}\right)^{\frac{1}{4}}}
\end{equation}
$T$ represents the total size of the sample (in this case, number of days). ADF tests return a test statistic $t$ that lies on a t-distribution with $df$ = $T - 1$. The resulting p-value of this analysis determines whether to reject the null hypothesis in favor of the alternative (equation \ref{eq:NullAltADF}), 

\begin{equation}\label{eq:NullAltADF}
{H_0: {\rho = 1} \ \ \ \ \ \ \ \ \ \ \ \ \ \ H_a: \rho < 1 }
\end{equation}

where $\rho$ represents the root. As implied by the name, $\rho$ = 1 signifies a unit root, while $\rho < 1$ indicates stationarity. Conversely, if one were testing for an explosive root, $H_a: \rho > 1$ would be used.

\bigbreak
\textbf{Lo-Mackinlay (1988) Variance Ratio (VR) Tests:} Variance ratio tests, conceptually outlined in \textit{Econometric Examination of Stochastic Price Trends}, are implemented by first finding the variance ratio over $k$-periods \cite{charles2009variance} with the calculation of

\begin{align}\label{eq:VarRat}
{Var(k) = \frac{Var(x_t +x_{t−1} + ··· +x_{t−k+1})/k}{Var(x_t)}} \\
{= 1 + 2\sum_{i=1}^{k - 1}(\frac{k - i}{k})\rho_i}
\end{align}

``where $\rho_i$ is the i-th lag autocorrelation coefficient of $x_t$" \cite{charles2009variance}. Using this logic for individual variance ratio calculation, the test statistic is calculated under the assumption that individual metrics are not $i.i.d$. The robust test statistic $M_2$ is denoted in equation 12.

\begin{equation}\label{eq:VarRatTstat}
{M_2(k) = \frac{VR(x;k) - 1}{\phi^* k^{(\frac{1}{2})}}}
\end{equation}
The coefficient $\phi^*$ represents the asymptotic variance under heteroskedastic conditions and is examined in the two formulas below \cite{charles2009variance} under the assumption $V(k) = 1$. 

\FloatBarrier
\begin{figure}[!ht]
\label{fig:RawPerf}
\centering
\includegraphics[width=1\linewidth]{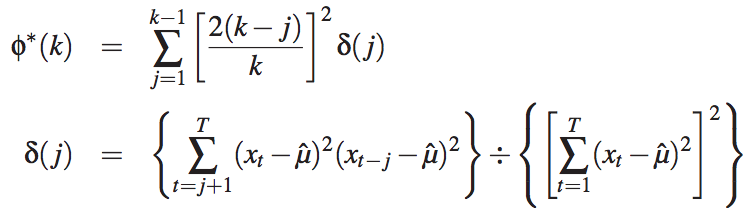}
\end{figure}
\FloatBarrier

In all tests conducted, we calculate $M_2(k)$ for daily intervals with the estimated mean $\hat{\mu}$. Daily intervals were chosen such that variance ratios corresponded to daily trading intervals. Test statistics and p-values are stored for all securities, the null hypothesis of Variance Ratio tests is equivalent to equation \ref{eq:NullAltADF}.

\subsection{Automating Fraud Detection Experiments}

To examine a variety of securities and models, a modular testing platform was created. Encapsulated in $Test$ objects, a specified stock-data sample and model run predict price movement for a specific security, as shown in Figure 1. Results are stored after each testing cycle in a $Logger$ object as a result log before the process repeats for a new security. If a trader is specified, stock predictions will be sent to an Automated Trader object which will execute and create a trade log documenting all portfolio activity. At the end of the experiment, a master log of all execution parameters, average results, and meta-data is sent to the $Logger$. 

\begin{figure}[h]
\label{fig:platform}
\centering\includegraphics[width=1\linewidth]{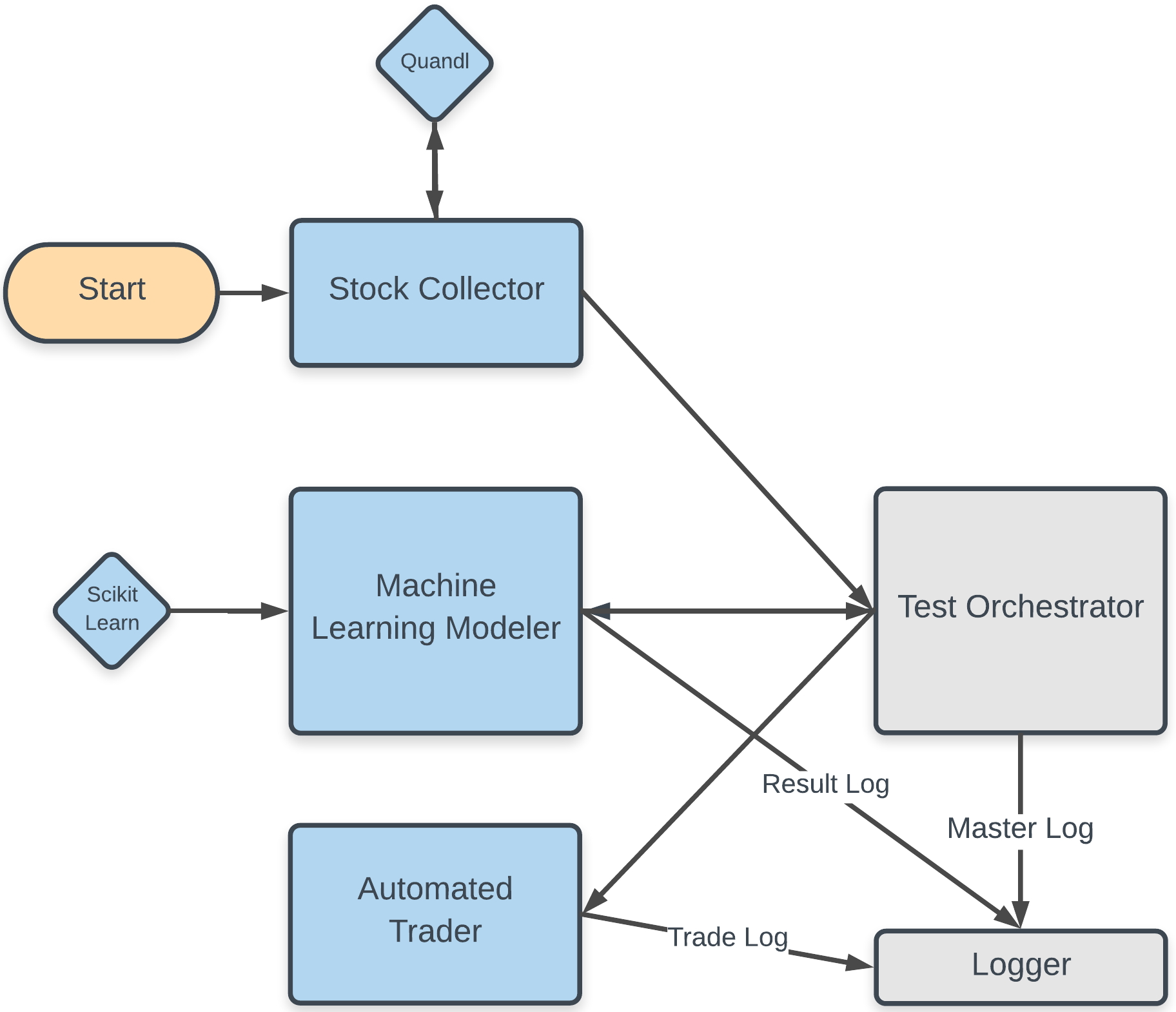}
\caption{Scalable automation platform for testing fraud detection algorithms}
\end{figure}

\textbf{Stock Data:} All executions begin with a command sent to the sampling engine that includes either a file name or stock ticker and a range of dates. The Stock Collector object will then either import stock prices from a flat file or download them via the Quandl API. Data first undergoes a quality assurance process that checks both for missing and ill-formatted data using outlier detection. If no issues are found, ten technical ratios are derived via the formulas in Figure 3. 

\begin{figure}[h]
\label{fig:TechnicalIndicators}
\centering\includegraphics[width=1\linewidth]{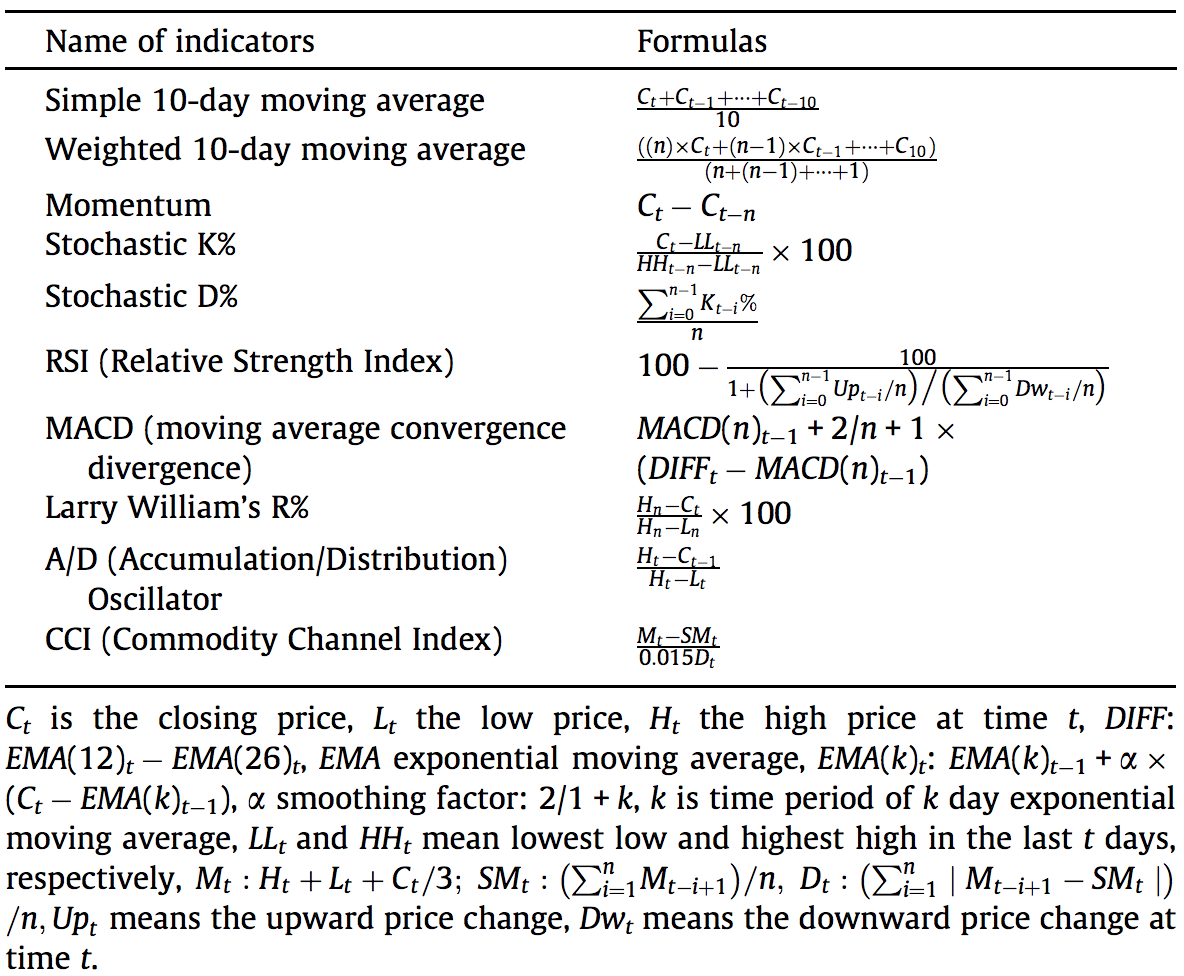}
\caption{Technical indicators as provided by \cite{Kara2011}}
\end{figure}

Moving averages and other indicators necessitate a loss of data
. The number of empty rows is counted and stored before records are removed. Daily price momentum
is also shifted to the previous day. As mentioned in \textit{Stock and Stock Index Data Sources}, technical ratios are utilized to predict price movement the following day. Therefore, by shifting price momentum back, a single row includes the technical indicators of that day as well as the price momentum 
of the following. Finally, the refined stock dataset is passed to the Machine Learning Modeler engine.

\bigbreak
\textbf{Modeling:} Provided a training sample, the Machine Learning Modeler begins predicting price movement. Just as a hedge fund will consistently re-train its system as new data arose, the algorithm implemented for this study tests a single day, then incorporates its prediction and results back into the model. This process repeats daily so all price information is considered.  Implemented with Sci-kit Learn, the modeler tests Logistic Regression, Support Vector Machines, Random Forest, K-Nearest Neighbors, and Gaussian Naive Bayes classifiers. Each model is trained and tested on the same range of data
, with performance metrics logged ad-hoc. 

All models are tested from 2008 to 2018 (~2520 days) to align findings with those of \cite{Patel2015} and \cite{Kara2011}. They are also recursive. With the initial training dataset, the platform will train the model but only predict the first record. The results if this execution are then fed back into the training dataset, and the modeler sends all of the stock data, now with added predictions, to an Automated Trader object.

\bigbreak

\textbf{Trading:} The Automated Trader determines buy, sell, and hold actions for a security based on a combination of momentum prediction and liquidity. If the trader is in a state of liquidity
and receives an upward prediction, it will issue a buy command and purchase as much of the security in question as possible
. Now that the trader is not in a state of liquidity, any further buy orders will be converted to holds until a sell order is executed. The same concept applies to repeated sell predictions, and all traders begin with \$100,000 of principal.

After buy, sell, and hold signals are assigned to each day, portfolio returns are calculated across the whole period. Transaction costs of \$4.95 are applied to all trades, aligning with the costs of Fidelity Brokerage, and factored in to total returns. Excess cash is invested at the current risk free rate, which has been discounted to daily returns. These two sums -- excess cash and equity assets -- sum to the total portfolio value and determine portfolio return. 

Portfolio return $R_p$ is employed to calculate several ancillary metrics, shown in Table 2. $R_m$ represents the return of the market benchmark, and $R_f$ is the risk free return of U.S. Treasury Bills. Regarding Table 2, $\sigma_p$ denotes portfolio volatility while  $\sigma_d$ denotes downside deviation.
.
\vspace{-2mm}
\FloatBarrier
\begin{table}[!ht]
    \centering
    \caption{Portfolio Performance Metrics}
    \begin{tabular}{
    	l
        l
        }
        \toprule
		{Metric} & {Formula}\\
        \midrule
		Alpha ($\alpha$) & $R_p - [R_f + (R_m - R_F)\beta]$\\[0.2cm]
        Beta ($\beta$) & $\frac{Cov(R_p, R_m)}{Var(R_m)}$\\[0.2cm]
        Volatility ($\sigma$) & $\sqrt{\frac{\sum_{i = 1}^{N}| R_i - \bar{R_p}| ^ 2}{N}}$\\[0.2cm]
        Sharpe & $\frac{\bar{R_p} - R_f}{\sigma_p}$\\[0.2cm]
        Sortino & $\frac{\bar{R_p} - R_f}{\sigma_d}$\\[0.2cm]
        \bottomrule
    \end{tabular}
\end{table}
\FloatBarrier
\vspace{-3mm}

\subsection{Implementing Stock Price Momentum Classifiers}

Five supervised learning models were chosen for predicting stock price momentum, and are outlined in the sections below. These were all chosen for their precedence in being applied to generate above-market returns through price movement predictions
. Hyperparameters for classifiers are outlined in Table 3.

\bigbreak
\textbf{Logistic Regression (LOG):} One of the most commonly implemented binary classification analyses \cite{hosmer2013applied}, Logistic Regression is a discriminant Bayesian model that approaches binary classification through direct calculation of Bayesian posterior $P(y|x)$ of the joint probability distribution $P(x,y)$ \cite{ng2002discriminative}. Logistic regressions assume little correlation (if any) exists between predictors in the feature space, as well as that few outliers and minimal skew are present. Despite this, many studies still implement Logistic Regression to predict price movements, and purport to have found success \cite{dutta2015prediction}. Prevailing research also finds discriminative calculation -- even without computational or data quality considerations -- to be generally preferred over generative models \cite{ng2002discriminative}. The empirical findings of this model will be compared with Gaussian Naive Bayes predictors to test this hypothesis.

Logistic regressions are tuned in this study by manipulating the regularization function and the regularization strength $C$.

\FloatBarrier
\begin{equation}\label{eq:LAD}
{L1 ={argmin_w\sum_{i = 1}^n{\Bigg[y_i -  \sum_{j=0}^m{w_j x_{ij}\Bigg]^2 + \lambda\sum_{j=0}^m{|w_j|}}}}}
\end{equation}

\begin{equation}\label{eq:LSE}
{L2 ={argmin_w\sum_{i = 1}^n{\Bigg[y_i -  \sum_{j=0}^m{w_j x_{ij}\Bigg]^2 + \lambda\sum_{j=0}^m{w_j^2}}}}}
\end{equation}
\FloatBarrier 

In machine learning, $L1$ loss is also known as Least Absolute Deviations (LAD) and included as part of regularization functions that classifiers seek to minimize. $L2$ is similar to $L1$ loss with the exception that it seeks to minimize the Least Squares Error (LSE).

Both of these regularization functions are categorized for weights $w$, output label $y$, and prediction $x$ in equations \ref{eq:LAD} and \ref{eq:LSE}. $C$, another parameter tuned for Logistic Regression, is a term for inverse regularization strength $\left(\frac{1}{\lambda}\right)$. This characterizes how harshly a model's complexity should be penalized during training. Classifiers will add terms to better fit to a dataset, but too much tuning causes over-fitting in the model and prevents it from generalizing trends to a population.

\bigbreak
\textbf{Support Vector Machine (SVM):} Vapnik (1995) \cite{Vapnik1995} first proposed Support Vector Machines (SVM) for pattern recognition by categorizing the problem as fitting an optimal separating hyperplane in $\mathbb{R}^n$ feature space, where $n$ is the number of features. By treating observations as $support \ vectors$, Vapnik characterizes the classifier as a Lagrangian optimization of the form:

\begin{equation}\label{eq:SVM}
{y(x) ={sign\Bigg[\sum_{i = 1}^N{\alpha_k y_k \Psi(\vec{x}, \vec{x}_k) + b \Bigg]}}}
\end{equation}
``for a training set of $N$ data points $\{y_k, \vec{x}_k\}_{k=1}^N$  where $\vec{x}_k \in \mathbb{R}^n$ is the $k^{th}$ input pattern and $y_k \in \mathbb{R}$ is the $k^{th}$ output pattern." \cite{suykens1999least}. $\alpha_k$ are positive, real constants, as is $b$. Notably, SVM makes use of kernel function $\Psi$, which transforms data for optimal for hyperplane separation. Since Vapnik discovered Support Vector Machine's efficacy at pattern recognition, variant classifiers have become common due to the flexibility afforded by kernel functions.  For example, Suykens (1999) \cite{suykens1999least} determined a popular method for implementing SVM with least squares optimization. Kernel research continues to derive new applications for different disciplines. 

\FloatBarrier
\begin{table*}[!ht]
    \centering
    \caption{Machine Learning Parameters by Model}
    \begin{tabular}{
    	l
        l
        l
        l
        l
        l
        l
        l
        l
        }
        \toprule
        \multicolumn{1}{c}{} &
        \multicolumn{7}{c}{Parameters}\\
        \cmidrule(lr){2-8} 
{Model}& {Kernel} & {Penalty}& $C$& {$\gamma$}& {$degree \ (d)$}& {Trees} & {$K$}\\
{Name} &  &  & &{$(rbf)$} & {$(poly)$} &{$(\#)$} &{$(\#)$}\\
        \midrule
		SVM& rbf, poly &&0.5, 1, 5 & auto, 1, 4& 1, 2, 3& & &\\
        LOG& & l1, l2  & 0.01, 1, 5, 10, 50, 100 & &\\
        RF &  &&&&& 20, 40 ... 100\\
        KNN &  &&&&  &&20, 40 ... 140\\
        \bottomrule
    \end{tabular}
\end{table*}
\FloatBarrier

Indeed, Support Vector Machines implement a variety of different kernels depending on the application. This study examines two of the most popular kernels, the radial basis function (rbf) and the polynomial (poly) kernel.

\begin{align*}\label{eq:SVMKernel}
{\textbf{rbf}: \ \ \ \ \ \ \ \ \ \ \ \ \ \ \ \ \ \ \ \ \ \ \ \ \ \ \ \ \ \Psi(\vec{x},\vec{x}_k) = e^{-\gamma{||\vec{x} - \vec{x}_k||^2}}} \\
{\textbf{poly}: \ \ \ \ \ \ \ \ \ \ \ \ \ \ \ \ \ \ \ \ \ \ \ \ \ \ \Psi(\vec{x},\vec{x}_k) = (\vec{x}^T\vec{x}_k + r) ^d}
\end{align*}

For RBF functions,  $\gamma$ is a free parameter that tunes the significance of a single training sample \cite{pedregosa2011scikit}. Additionally, $C$ trades off misclassification of training examples against simplicity of the decision surface (not shown in equations). On the other hand, $d$ denotes the degree of the polynomial and r is a constant that trades off the weighting of higher and lower order terms in the polynomial \cite{shashua2009introduction}. $r$ is not manipulated and left as the default of zero. Other kernels exist, but RBF and POLY were implemented by \cite{Patel2015} and \cite{Kara2011}. Since these studies each examined non-U.S.
financial markets, we wish to compare our performance in American markets while controlling for as many extraneous factors as possible.

\bigbreak
\textbf{Random Forests (RF):} Breiman (2001) \cite{Breiman2001} defines Random Forests as ``a combination of tree predictors such that each tree depends on the values of a random vector." An ensemble algorithm, Random Forest classifiers derive their efficiency from a sufficiently large number of decision trees. Each decision tree in a group may suffer from high generalization error and overfitting. However, if taken together as a random forest voting system, the ensemble has been proven to produce a "limiting ... generalization error" \cite{Breiman2001}. 


\bigbreak
\textbf{Gaussian Naive Bayes (GNB):} As opposed to the discriminant Bayesian implementation employed by Logistic Regression \cite{ng2002discriminative}, Gaussian Naive Bayes classification empirically generates
the joint probability distribution $P(x,y)$ and then uses Bayes Theorem to calculate $P(y|x)$. Favored for their simple and efficient implementation, Bayesian models consistently perform well across a variety of applications \cite{zhang2004optimality}. Outlined by \cite{Lewis1998}, Gaussian Naive Bayesian models assume features are independent and model a normal distribution. 

\begin{equation}\label{eq:NormBayesCalc}
{f(x \ | \ \mu, \sigma) = \frac{1}{\sqrt{2\pi}{\sigma}}{e}^{{-\frac{(x - \mu)^2}{2\sigma^2}}}}
\end{equation}
By using an empirically discovered $\hat{\mu}$ and $\hat{\sigma}$ for each feature, a Gaussian probability distribution can be derived, as shown in equation \ref{eq:NormBayesCalc}. This information is then be fed into record classification in equation \ref{eq:BayesianCalc}, where $P(A|B)$ is equal to the product of the probabilities of each feature $B_i$ belonging to the normal distribution of that feature $N(\mu_{B_i}, \sigma_{B_i})$ derived from training data.
\begin{equation}\label{eq:BayesianCalc}
{P(A | B) = P(A)  \prod_{i=1}^{n}P(B_i | A)}
\end{equation}

\bigbreak
\textbf{K-Nearest Neighbors (KNN):} The control classifier, K-Nearest Neighbors (KNN) is a rudimentary machine learning classifier. Though highly dependent on the dataset and application, KNN can be an effective means of prediction. A test prediction is generated by finding the $K$ most similar records in the training data, and returning the most common class label in the set. Similarity $d$ of a training record $x$ relative to test record $y$ is often determined using the Euclidean distance (\ref{eq:KNN}) function, defined for a dataset with $n$ features as

\begin{equation}\label{eq:KNN}
{d(x) = \sum_{i=1}^n{\sqrt{(x_i - y_i)^2}}}
\end{equation}

where $x_i$ is the $i^{th}$feature of the training record. KNN has shown to be an effective algorithm for some classification problems, but is known \cite{imandoust2013application} to struggle in many applications. Since predictions are derived purely from records seen in the training dataset,  KNN struggles with economic problems involving sparse feature spaces (e.g. continuous price data).

%

\section{Empirical Findings}
\label{S:8}

\subsection{Econometric Findings Indicate Possible Predictability}

Test statistics from Augmented Dickey-Fuller tests unanimously reject the null hypothesis of a unit root in favor of stationarity. The Akaike Information Criteria assigned between 25 and 28 lags for each test, and p-values were well below 1\% significance. If the maximum lag is allowed to double from what was previously determined by \cite{schwert1989tests}, all test statistics translate to higher values, but even then the majority of tests still rest below a 1\% significance value. These results, shown in Figure 4, present compelling evidence that price movements for 100 stocks from 1998 to 2008 are not entirely random.

With that said, maximum lag levels are difficult to set appropriately. Adding lags until statistic significance no longer exists may not necessarily be an appropriate approach \cite{dickey1979distribution}. It also can be conducted in different ways, including the Akaike and Bayesian Information Criteria. Regardless, setting lags is not the purpose of this analysis, and it will be assumed that the recommendations of previous literature \cite{schwert1989tests} are sound in this application.

\FloatBarrier
\begin{figure}[!hbt]
\label{fig:ADFTestStat}
\centering
\includegraphics[width=1\linewidth]{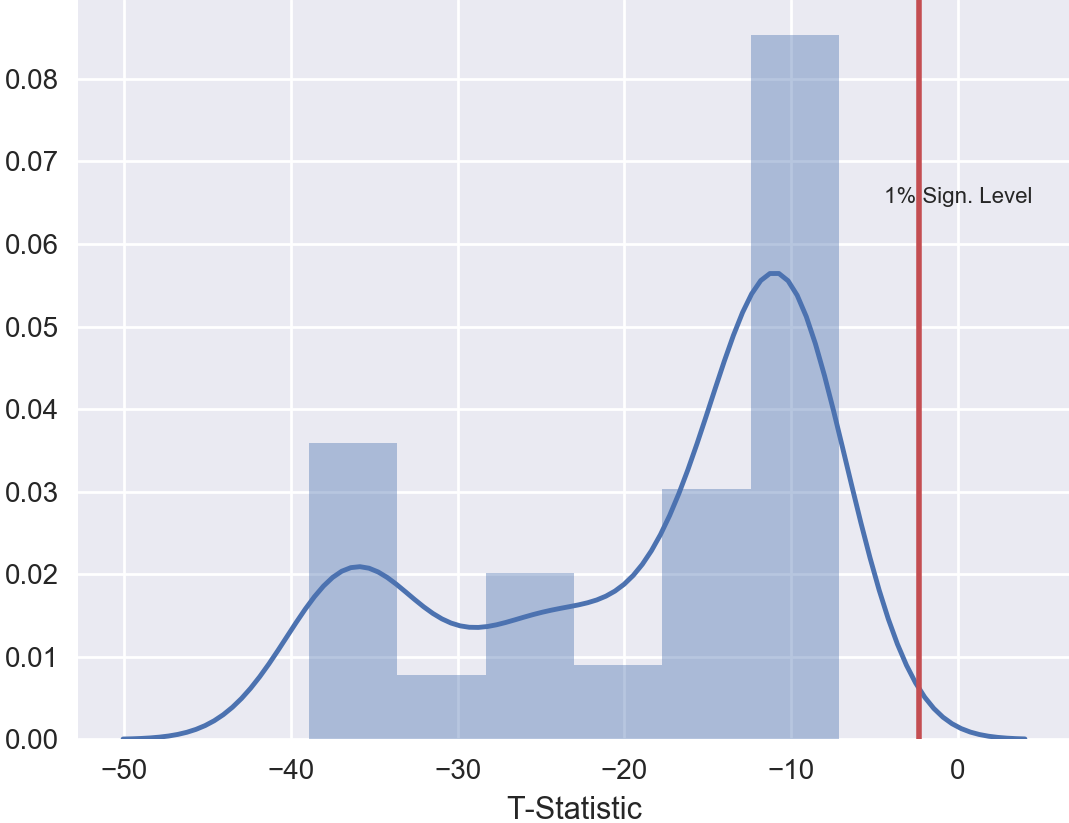}
\caption{Relative frequency histogram of T-statistics for Augmented Dickey-Fuller tests of 100 randomly selected S\&P 500 securities over 10 years.}
\end{figure}
\FloatBarrier

\FloatBarrier
\begin{figure}[h]
\label{fig:LoMac}
\centering
\includegraphics[width=1\linewidth]{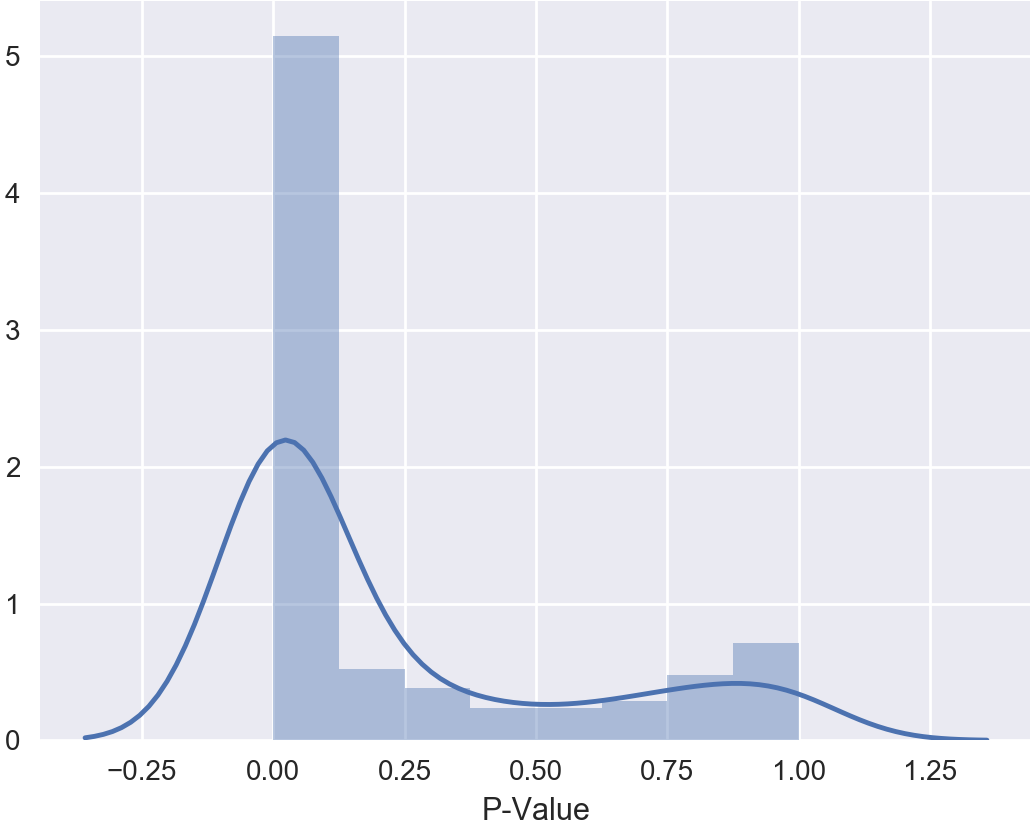}
\caption{Relative frequency histogram of P-values for Lo-Mackinlay Variance Ratio tests of 100 randomly selected S\&P 500 securities over 10 years.}
\end{figure}
\FloatBarrier

To better gauge the validity of Augmented Dickey Fuller findings, Variance Ratio tests for the same securities were conducted over the same time period. VR test results depict a similar -- though less stark -- verification of ADF findings. Shown in Figure 5,  93\% and 70\%  of p-values rest below significance levels of 5\% and 1\%, respectively. Repeated testing with different securities did not yield a unanimous verdict on the Random Walk Hypothesis, but does reinforce the notion that some level of price predictability existed in the past 20 years.

Unfortunately, Augmented Dickey-Fuller and Variance Ratio tests do nothing more than identify stationarity in general. No deductions can be made about the characterization or duration of trending in financial markets. Conclusions drawn by \cite{LoMacNonRand} articulate that a smaller variance ratio than expected can be seen as evidence of mean reversion
. Their deduction may indeed be correct, but variance ratios can not be definitively categorized as evidence for mean-reversion without further information. Rejecting a null hypothesis is not equivalent to accepting an alternative. It simply proffers the opportunity for further analysis.

Indeed, the following sections expound on the trends witnessed when machine learning algorithms are applied to predict price movements and inform trading decisions. Though these experiments do not touch on mean reversion directly, they test the ability of supervised learning algorithms to discern price movement trends.

\subsection{Price Predictions Align with Random Chance}

Across all machine learning models tested
, prediction accuracy did not statistically deviate from approximately random chance. Shown in the horizontal histogram of Figure 6, predictive accuracy observations were approximately normally distributed about a mean of 52\% with a slight left skew. Average accuracy may rest above 50\% because upward momentum is slightly more likely long term, though this hypothesis is unverified.  Trading performance (Alpha) above the risk free return $R_f$ possessed a similar distribution centered on returns of zero. 

Empirical distributions found for accuracy and alpha both align with the weak-form Efficient Market Hypothesis. The active return gained on an investment, Alpha represents the returns acquired above and beyond the market benchmark. weak-form market efficiency implies that beating the market via the exploitation of price trends is impossible. Indeed, with a plurality of models only able to achieve the market return or worse, trading results thus far seem to strongly affirm efficient market theories.

Notably, accuracy has a fairly strong, positive relationship with Alpha. It is expected that an algorithm effective at predicting price movement would generate a higher return. Even so, witnessing this trend empirically
verifies the notion that algorithmic trading platforms may be capable of success. The steep slope of returns against accuracy implies that a model with even 65\% accuracy over a long period of time could significantly outperform a market benchmark with our strategies. However, it is important to note that these strategies do take on significantly more risk.

\begin{figure}[!ht]
\label{fig:RawPerf}
\centering
\includegraphics[width=1\linewidth]{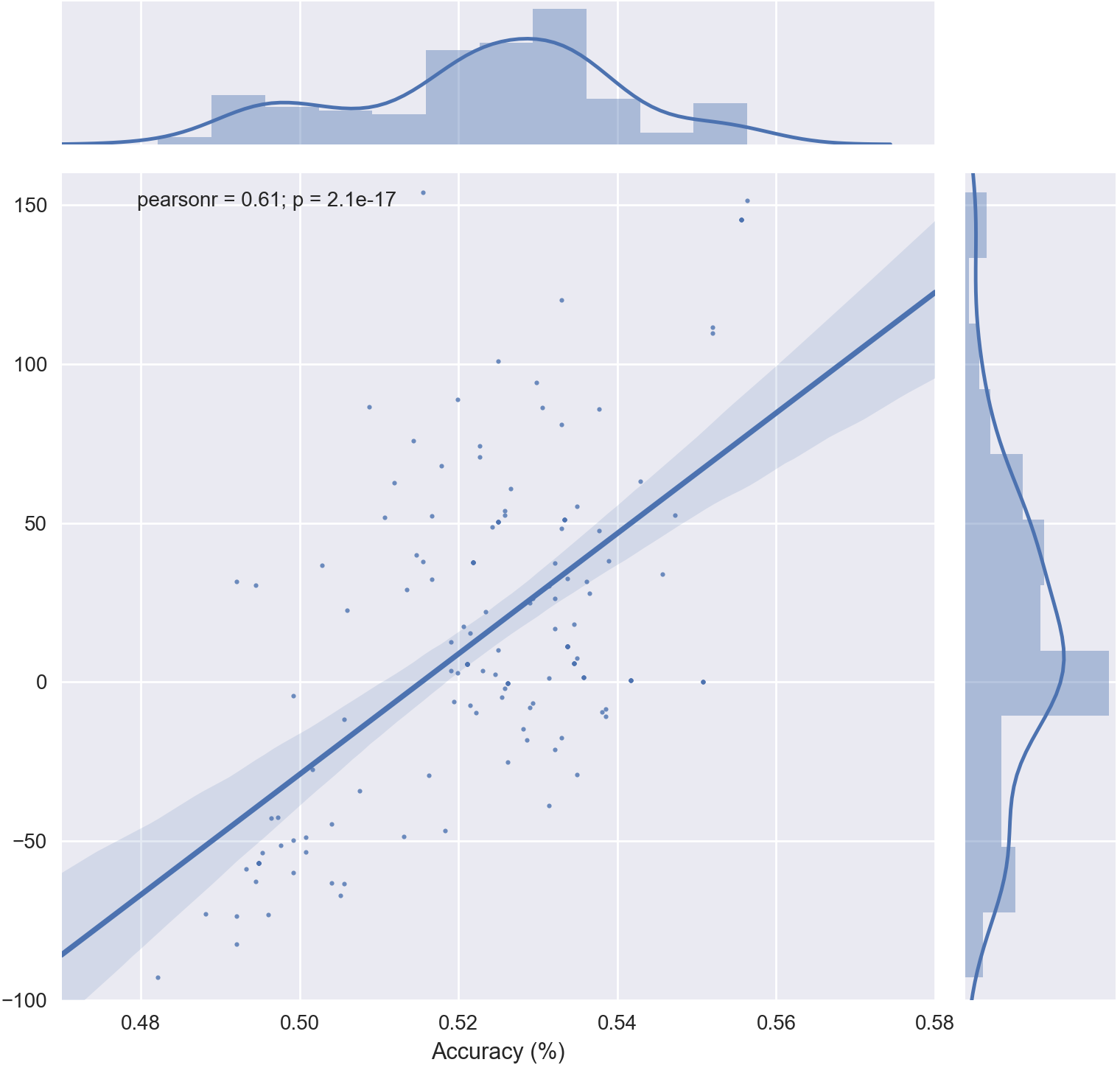}
\caption{Trend specific trading returns (Alpha) for all models, mapped as a function of accuracy. Alpha is represented as percent return over 10 years. Hue represents a 95\% confidence interval.}
\end{figure}

%

The positive relationship between Accuracy and Alpha was derived from less than 100 independent measurements. Due to the structure of the algorithmic trading model, all tests aside from RF, which naturally varies even when parameters are not changed, are static in their performance
. While the relationship between Alpha and Accuracy--as well as performance overall--aligns with efficient market theories and previous financial research, more testing is necessary to ascertain if our assertions are sound.

Accuracy overall appeared to revert to a mean of 50\%, but Figure 7 indicates some models consistently outperform others. Denoted by a specific color, each model's accuracy and Alpha performance data was regressed.  Across all four stock indices, no notable differences can be ascertained for the majority of models. However, the Logistic regression's average performance exceeded its peer algorithms in both accuracy and portfolio returns. 

\FloatBarrier
\begin{figure}[!hb]
\label{fig:AlphaAccByModel}
\centering
\includegraphics[width=1\linewidth]{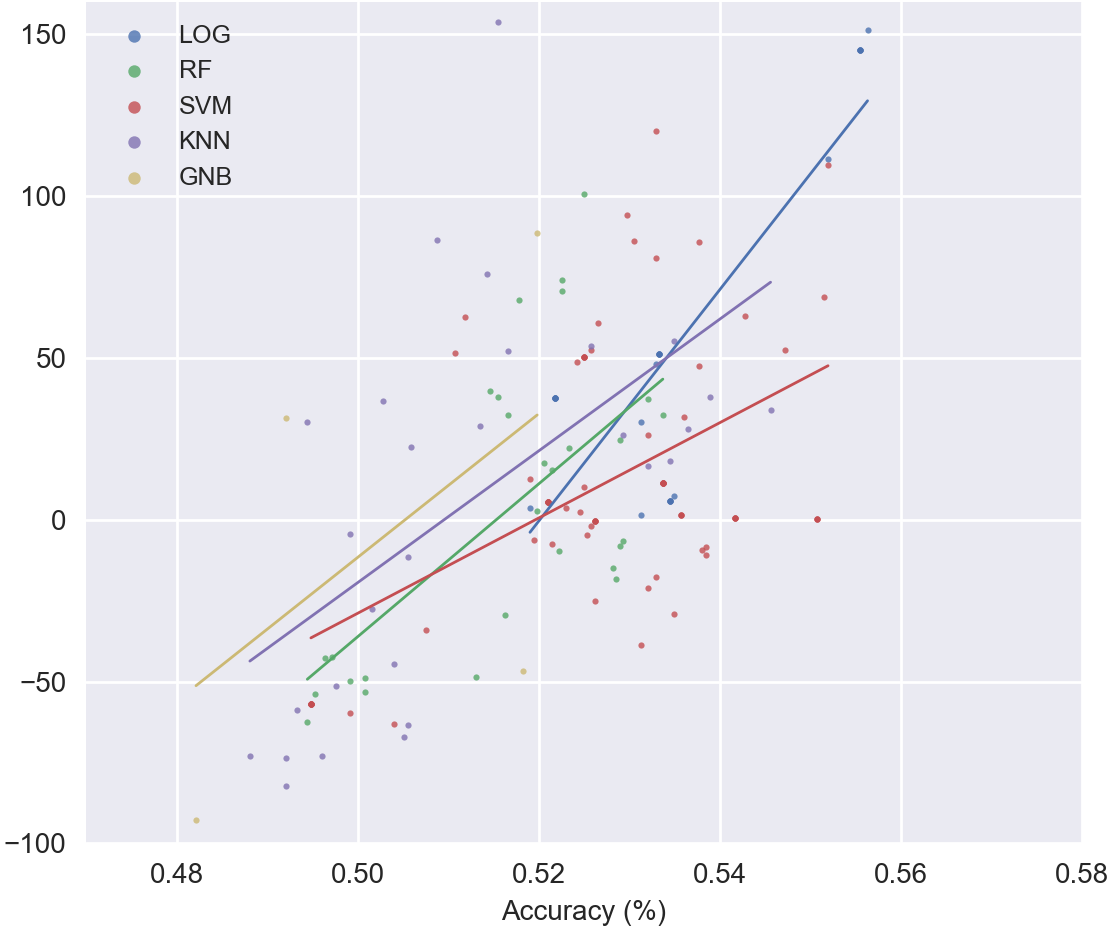}
\caption{Trend specific trading returns (Alpha) by model, mapped as a function of accuracy. Alpha is represented as percent return over 10 years.}
\end{figure}
\FloatBarrier

The cause of Logistic Regression's abnormal returns is unknown, and may be a manifestation of random chance.  Yet, when utilized to generate predictions for trading, it outperformed or matched the performance of the market for all four stock indices. Furthermore, it was the best performing algorithm in over 75\% of trade simulations run
. These findings also mirror the findings of \cite{ng2002discriminative}, who asserted the superiority of discriminative models for many classification problems. 

Consistently outperforming its market benchmark, Logistic Regression may have identified subtle market trends. Despite its inability to predict price movements with more than 56\% accuracy, Logistic Regression above-market returns skyrocketed to a maximum of 150\% over ten years. However, the validity and consistency with which Logistic Regression can achieve abnormal returns is questionable.

One interpretation of Logistic Regression trading is that it discerns strong price movements and correctly trades to exploit them. Such a strategy could easily gain abnormal returns without accurately predicting a majority of price movements. Another equally valid perspective is that the returns witnessed were random and anomalous, implying Logistic Regression is incapable of consistently outperforming the market if applied to other indices and securities. Independent testing, conducted and articulated in the following sections, supports the latter.

\FloatBarrier
\begin{figure}[!ht]
\label{fig:BetaAlphaRel}
\centering
\includegraphics[width=1\linewidth]{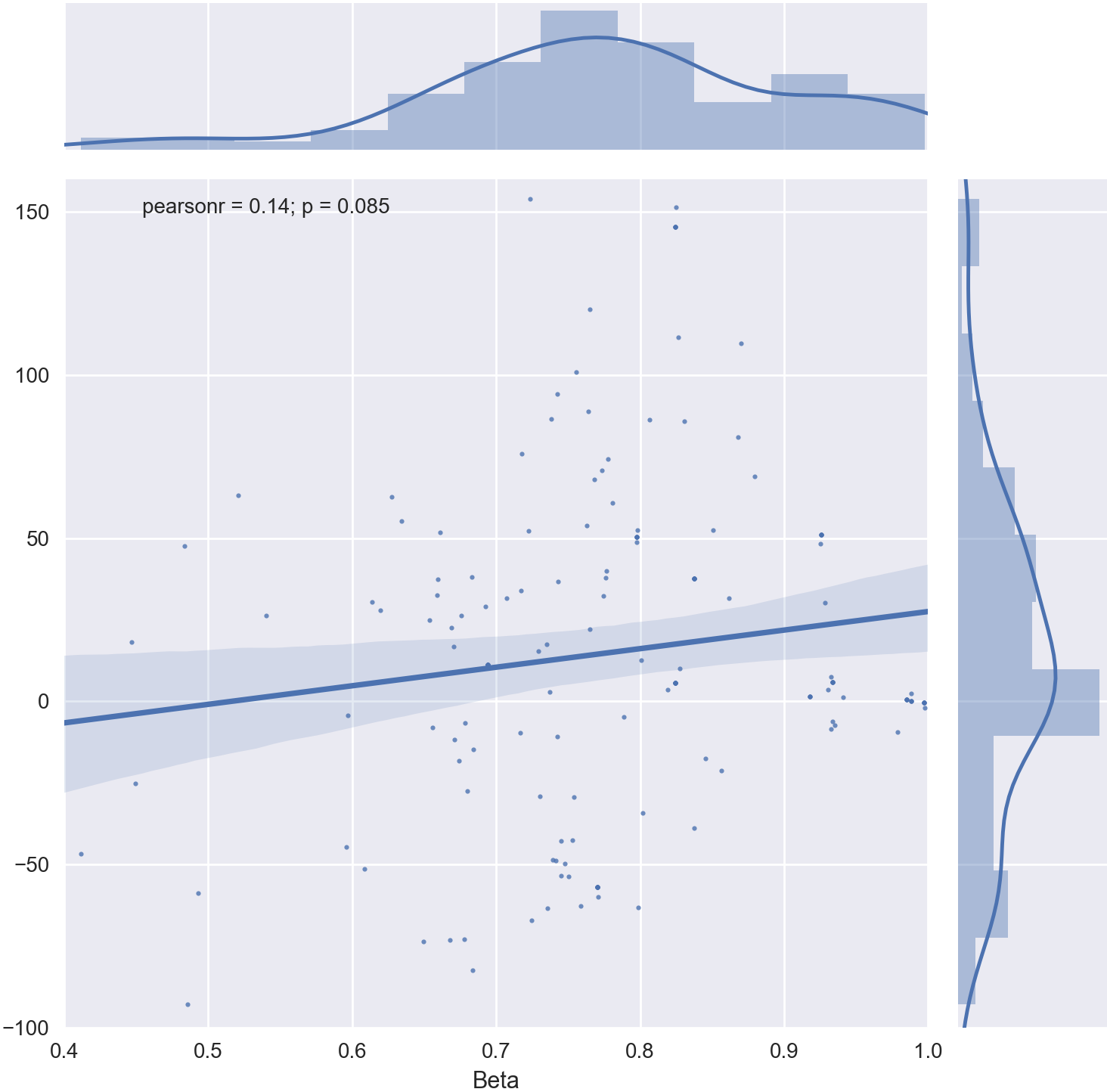}
\caption{Trend specific trading returns (Alpha) for all models, mapped as a function of Beta. Alpha is represented as percent return over 10 years. Hue represents a 95\% confidence interval.}
\end{figure}
\FloatBarrier

Trading performance bore little relationship with the Beta of the trading strategy (Figure 8). The empirical findings of \cite{FamaFrench2004} also find Beta to be uncorrelated with overall returns, though their analysis utilizes Beta as a proxy for the appeal and expected returns of a company's stock. A slight positive trend between Beta and Alpha is observed, but the lack of correlation increases the likelihood that the trend is the product of randomness. We maintain that any trend between Beta and Alpha would dissipate as additional experimental observations are added.

Additionally, Beta coefficients for all trading models are never negative and never exceed 1. Contextually, this means that changes in portfolio return are constrained to either mirror the movements of the market or move unrelated to it entirely. For many fund managers, a relatively low Beta is desired as it implies returns can be gained regardless of market volatility
. Positive Beta values also indicate trading algorithms generally do not move against the market benchmark. Both of these constraints in Beta are largely due to the structure of the algorithmic trading system.

Short selling is not programmed into the trading systems of this study. Therefore, the only states of a portfolio are vested in the market or liquid and receiving risk free returns. Even if the trading algorithm invested and sold at inopportune times, losses would not achieve a Beta lower than zero. Restrictions on Beta values above one are also attributed to the same design patterns.

\FloatBarrier
\begin{figure}[!ht]
\label{fig:VolAlpha}
\centering
\includegraphics[width=1\linewidth]{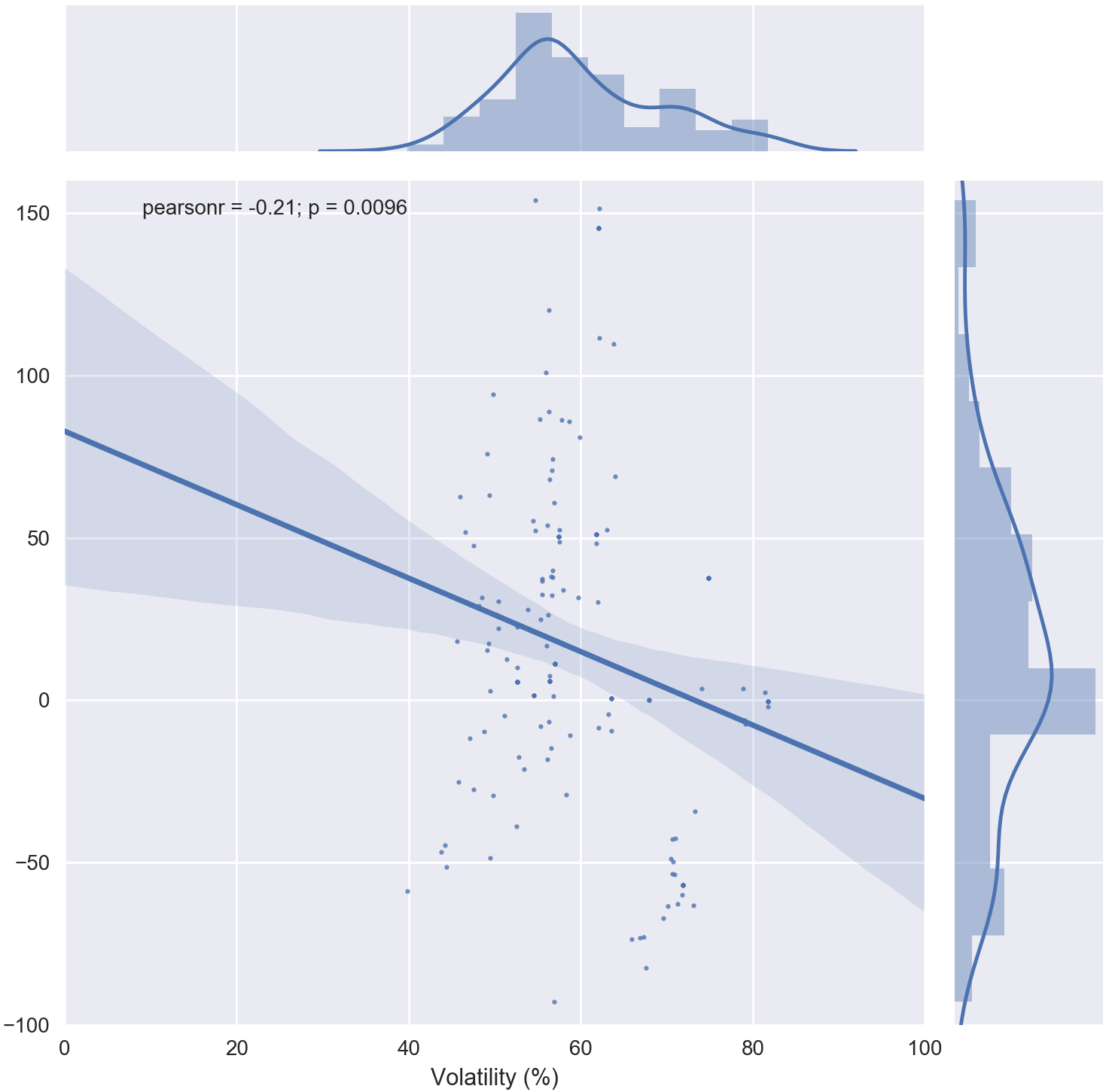}
\caption{Trend specific trading returns (Alpha) for all models, mapped as a function of Volatility. Alpha is represented as percent return over 10 years. Hue represents a 95\% confidence interval.}
\end{figure}
\FloatBarrier

Little information on the relationship between Alpha and Volatility can be gleaned from this study as little correlation exists. Similar to the trending seen in Figure 8, Figure 9 illustrates a slight negative relationship between Volatility and Alpha. High variation in data causes any conclusions on this relationship to be dubious. A collection of experiments separated from general findings 
all possess consistently poor performance and high volatility. If these outliers are removed, the negative relationship disappears. Additional experimentation is necessary to distinguish a Volatility-Alpha relationship for this algorithmic trading system.

\subsection{Trading Systems Generally Unable Outperform Market}

Aggregate trends outlined in the previous section are visualized in Figures 10 - 17 and grouped by index. Each index outlines trading performance for all models using either trending or non-trending data. The effects of non-trending data on performance are left for a later section, and focus here will be placed exclusively on trending data (Figures 10, 12, 14, and 16). Trading platform structure also induces trading performance trends to mirror the benchmark, and was expected. An algorithm that correctly differentiates between up and down price movements over time would mirror market trends while simultaneously exceeding market performance. 

\FloatBarrier
\begin{figure}[!ht]
\label{fig:RawPerf}
\centering
\includegraphics[width=1\linewidth]{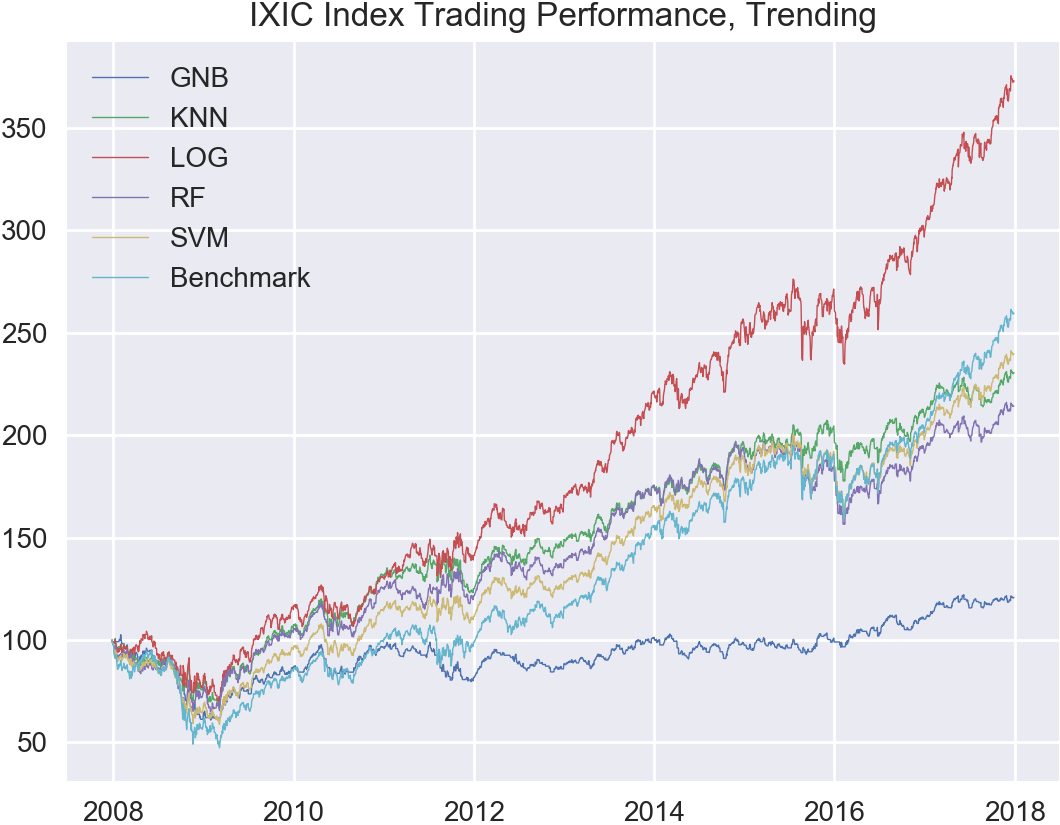}
\caption{IXIC 10 year trading performance for non-trending algorithms by model. Returns represented in USD - thousands. Benchmark represents buy-and-hold strategy.}
 \vspace*{\floatsep}
  \vspace*{\floatsep}
\includegraphics[width=1\linewidth]{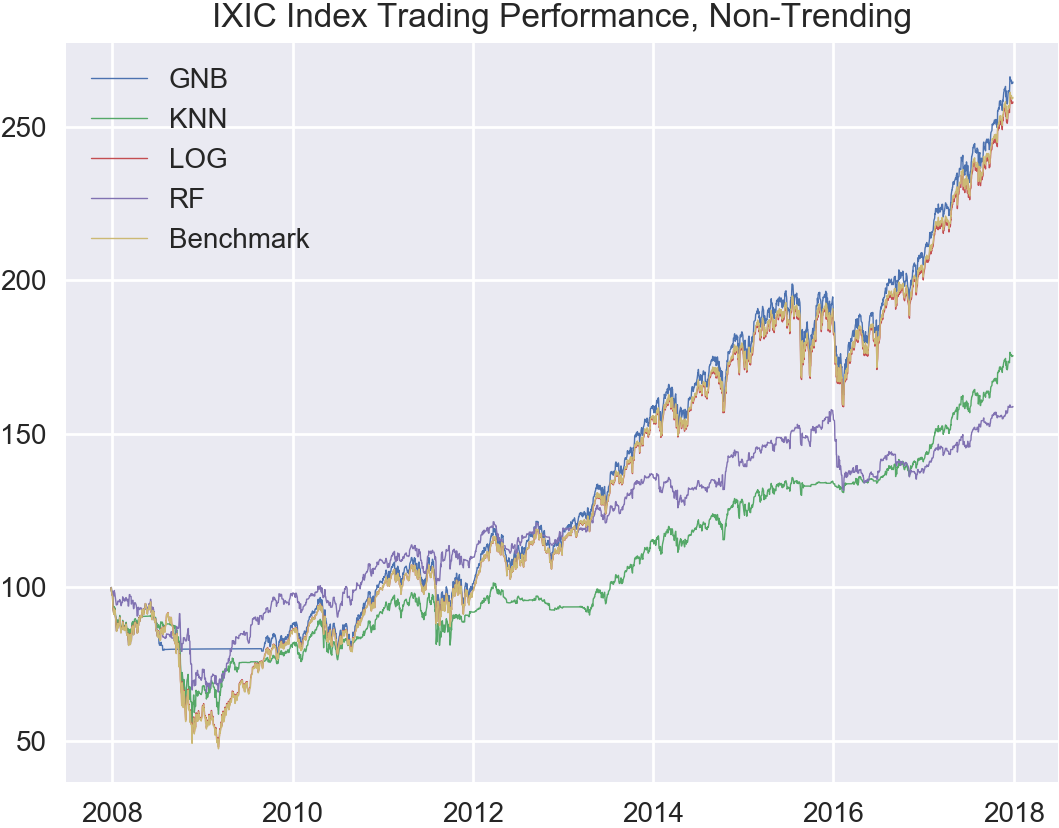}
\caption{IXIC 10 year trading performance for non-trending algorithms by model. Returns represented in USD - thousands. Benchmark represents buy-and-hold strategy.}
  
\end{figure}
\FloatBarrier

The performance of Logistic Regression in Figure 10 illustrates this point. Portfolio returns exceed the market at an increasing rate, and price shifts in the benchmark more profoundly affect Logistic Regression volatility. Other models do not exhibit this ability to identify and capitalize on volatility.


\FloatBarrier
\begin{figure}[!ht]
\label{fig:RawPerf}
\centering
\includegraphics[width=1\linewidth]{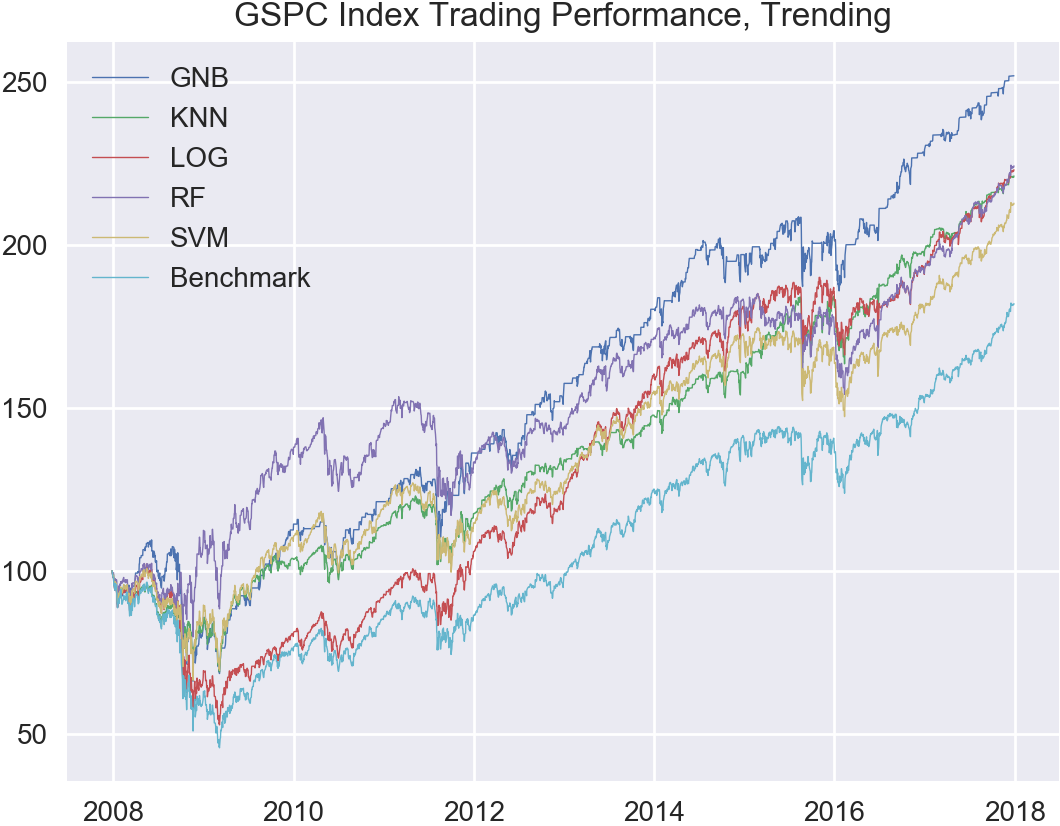}
\caption{GSPC 10 year trading performance for trending algorithms by model. Returns represented in USD - thousands. Benchmark represents buy-and-hold strategy.}
 \vspace*{\floatsep}
  \vspace*{\floatsep}
\includegraphics[width=1\linewidth]{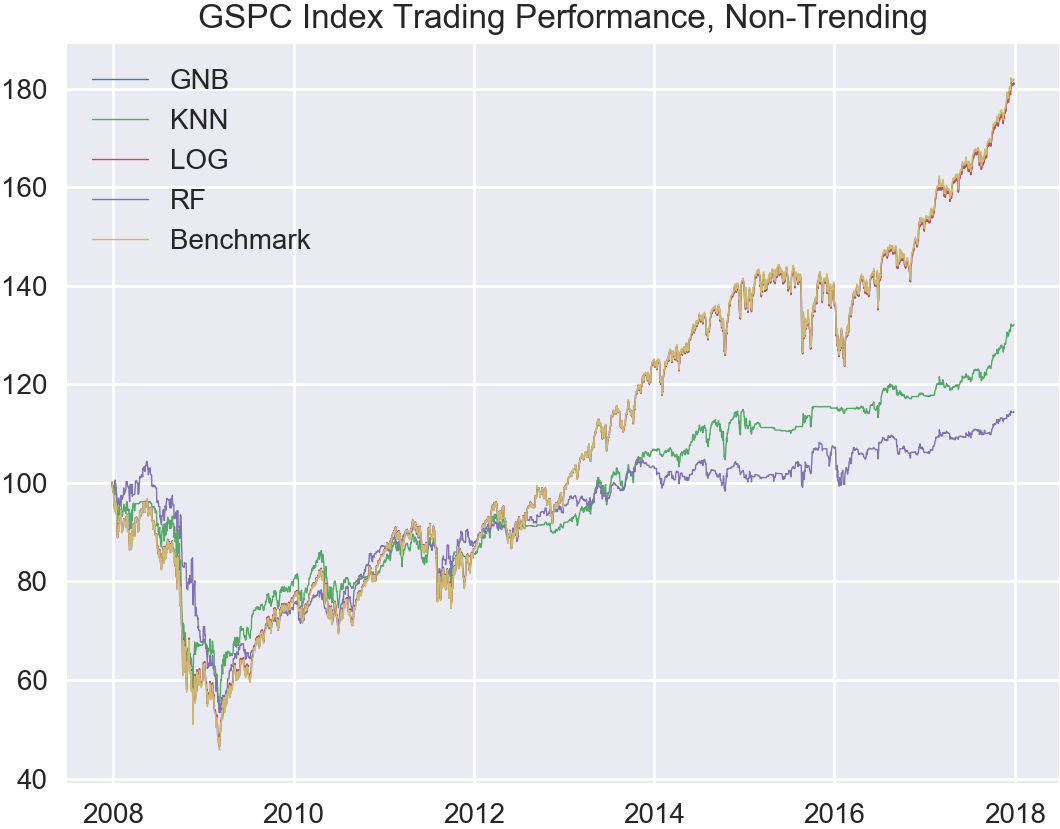}
\caption{GSPC 10 year trading performance for non-trending algorithms by model. Returns represented in USD - thousands. Benchmark represents buy-and-hold strategy.}
  
\end{figure}
\FloatBarrier

Logistic Regression completed an average of 360
trades over a ten year period, well under an average of 720 for RF and KNN, 600 for SVM, and 465 for GNB. Total transaction costs were much lower for Logistic Regression as a result, but likely not greatly influential on performance.

Conversely, LOG consistently garnered the highest accuracy of all models at 55.56\%, which greatly altered its 10-year trading performance. Trade log review indicates a tendency for LOG to execute trades shortly before periods of high volatility. The opposite can be said for trade results of Gaussian Naive Bayes classifiers.


\FloatBarrier
\begin{figure}[!ht]
\label{fig:RawPerf}
\centering
\includegraphics[width=1\linewidth]{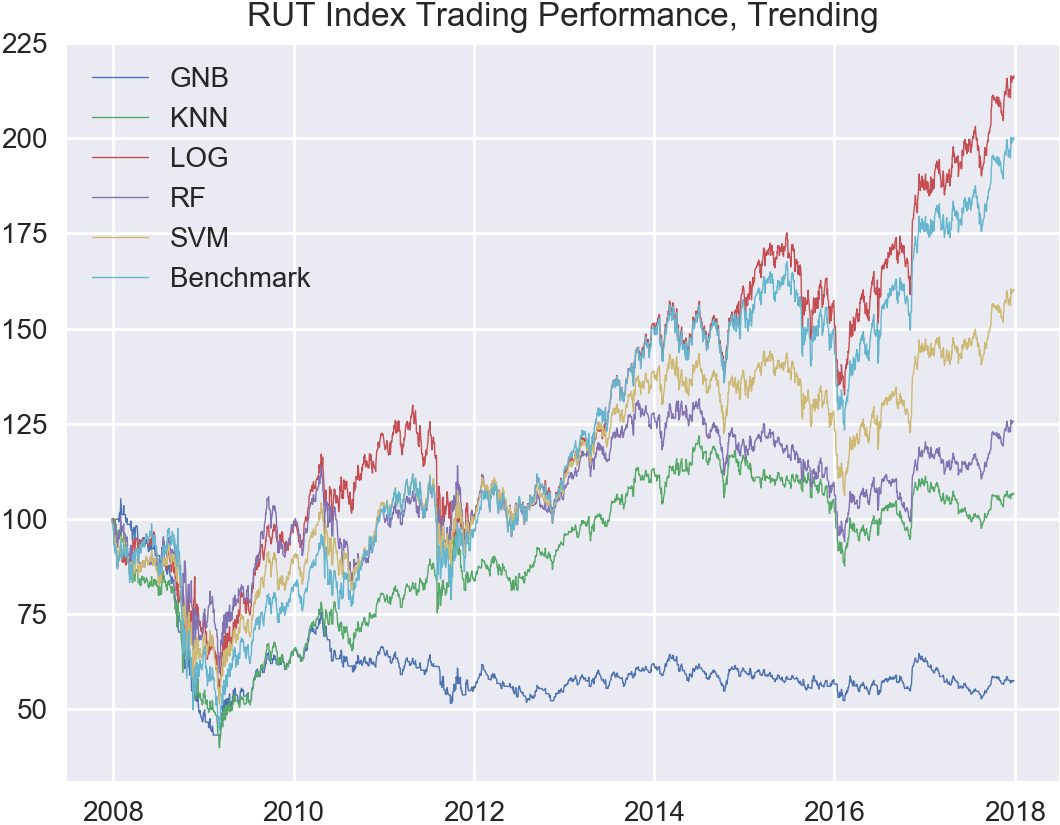}
\caption{RUT 10 year trading performance for trending algorithms by model. Returns represented in USD - thousands. Benchmark represents buy-and-hold strategy.}
 \vspace*{\floatsep}
  \vspace*{\floatsep}
\includegraphics[width=1\linewidth]{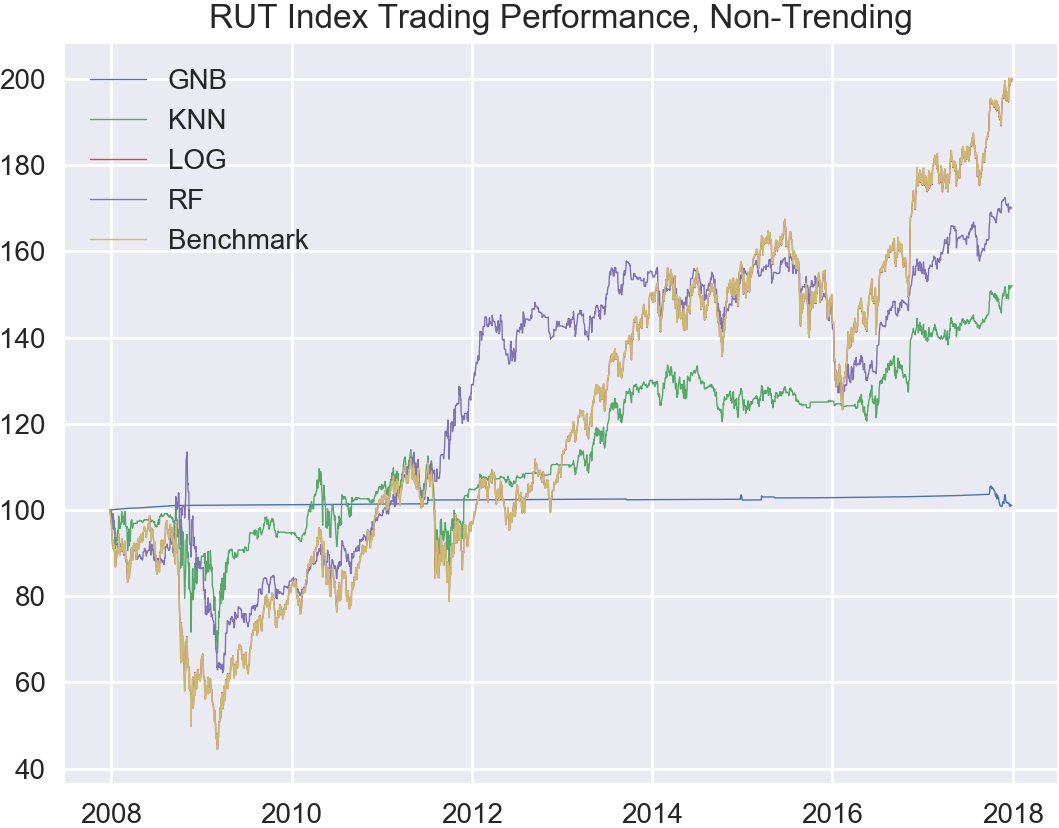}
\caption{RUT 10 year trading performance for non-trending algorithms by model. Returns represented in USD - thousands. Benchmark represents buy-and-hold strategy.}
  
\end{figure}
\FloatBarrier

Though most similar in statistical implementation and number of trades to LOG, GNB traded the worst out of any model for IXIC. Trades were often made sporadically, showing no relationship to price movement whatsoever. RF, KNN, and SVM models achieved near-market success. Holding periods were much shorter for these securities on average
. However, upon closer examination LOG and GNB often alternated frequent, near daily trades with holding periods of roughly two and a half weeks
. Other models tended to trade at more consistent intervals.


The cadence and number of trades remained consistent for each model as they were applied to different stock indices
. Surprisingly, Gaussian Naive Bayes classifiers garnered the highest return for GSPC and DJI but the lowest return for IXIC and RUT. Trading frequency and cadence is strikingly similar across all four indices. Trade logs found that in general, the distinction in performance was attributable to GNB trading IXIC and RUT into a losing position shortly before large price movements and holding the position until volatility dropped. By contrast, GNB consistently traded into and held winning positions in DJI and GSPC. 

\FloatBarrier
\begin{figure}[!ht]
\label{fig:RawPerf}
\centering
\includegraphics[width=1\linewidth]{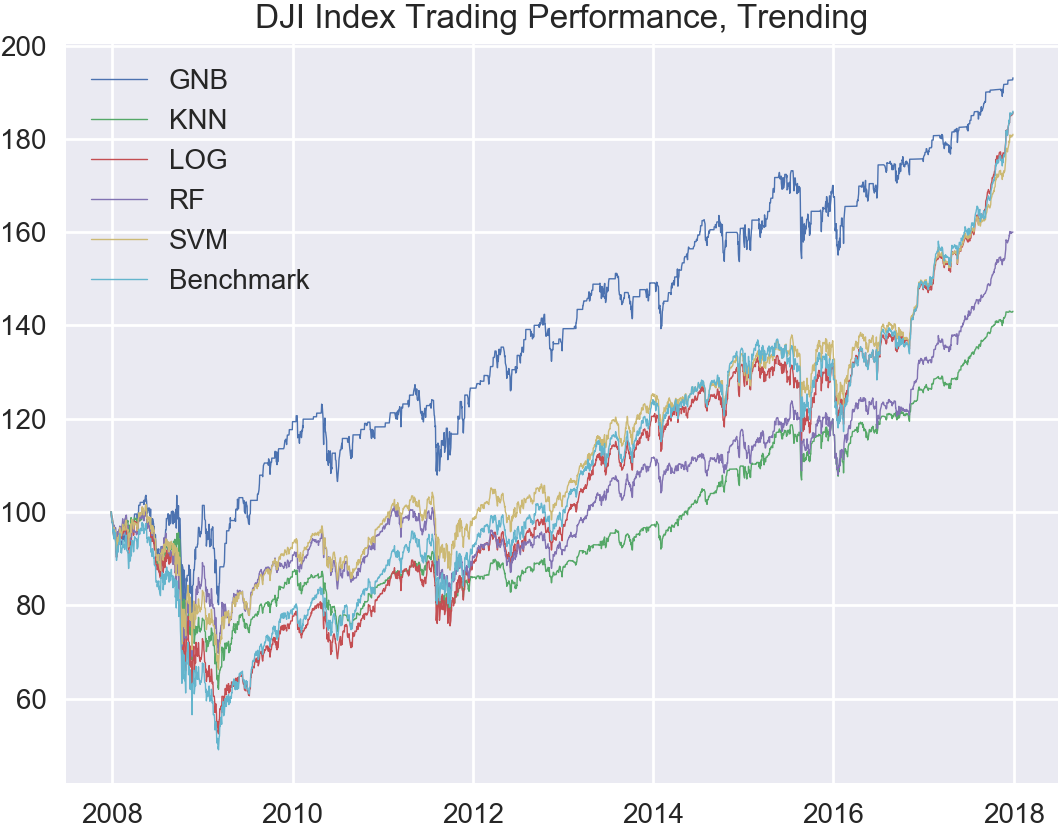}
\caption{DJI 10 year trading performance for trending algorithms by model. Returns represented in USD - thousands. Benchmark represents buy-and-hold strategy.}
 \vspace*{\floatsep}
  \vspace*{\floatsep}
\includegraphics[width=1\linewidth]{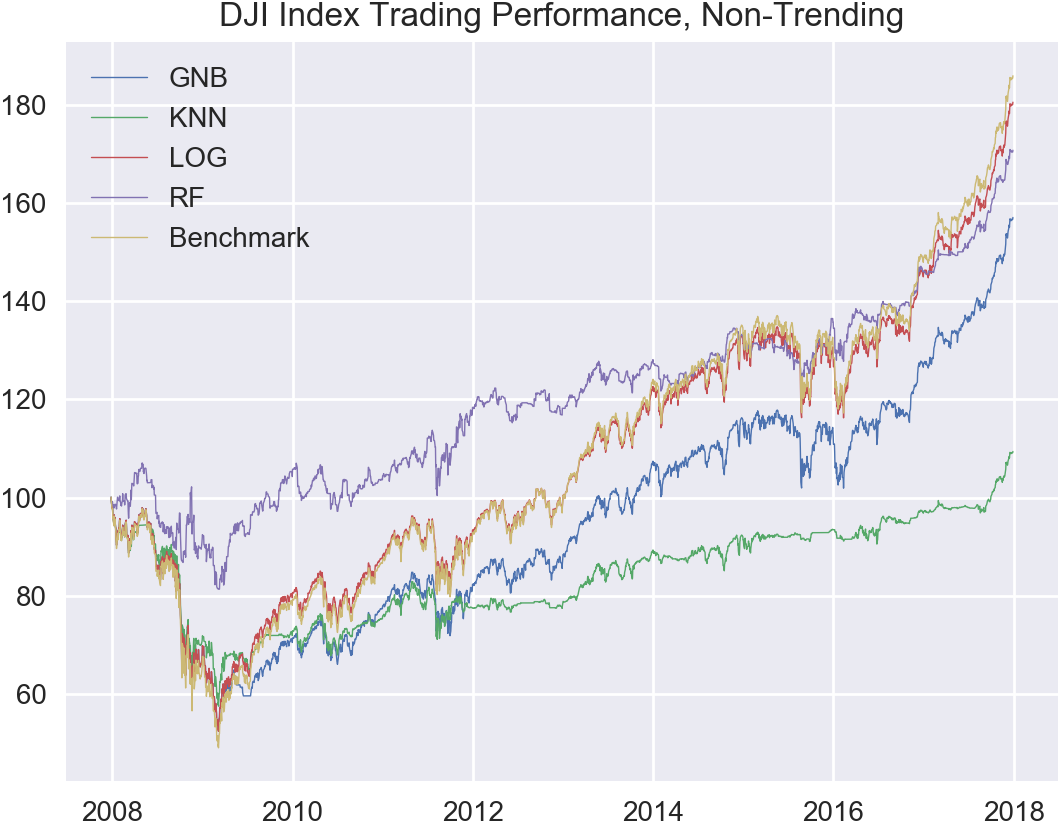}
\caption{IXIC 10 year trading performance for non-trending algorithms by model. Returns represented in USD - thousands. Benchmark represents buy-and-hold strategy.}
  
\end{figure}
\FloatBarrier

The origin of this simple but influential discrepancy in performance for GNB is not immediately clear. Variation in the consistently of GNB performance affirms the notion that these results are the product of random trading decisions. Generally, all models do not notably exhibit rules with which they use to trade, suggesting they have no predictive power in determining price changes.

Even if price momentum algorithms are picking up on trends, the volatility of the most successful models was significantly higher than the market benchmark. Logistic Regression's volatility over ten years was 65\% on average, while RF and KNN maintained volatility of 55\% and GNB 50\%. Additional risk in finance correlates to additional reward \cite{Fama1992}. Even if it could be proven that model predictions were identifying trends in the market, above-market returns can only refute efficient market theories if risk undertaken by the agent can be controlled for and minimized. 

Hedging with the strategies utilized by this study may propose a more convincing argument against weak-form EMH. Assuming that a model can be found that consistently outperforms the market, one could undertake small counter-positions in a separate index to assuage the risk of the portfolio while maintaining above-market returns. Logistic Regression appears to be the only model that may be able to consistently receive excess returns, and is more thoroughly tested on 100 randomly selected S\&P500 securities in the next section.


\subsection{Independent Verification Implies Performance Random}

\FloatBarrier
\begin{figure}[ht]
\label{fig:Rand100Acc}
\centering
\includegraphics[width=1\linewidth]{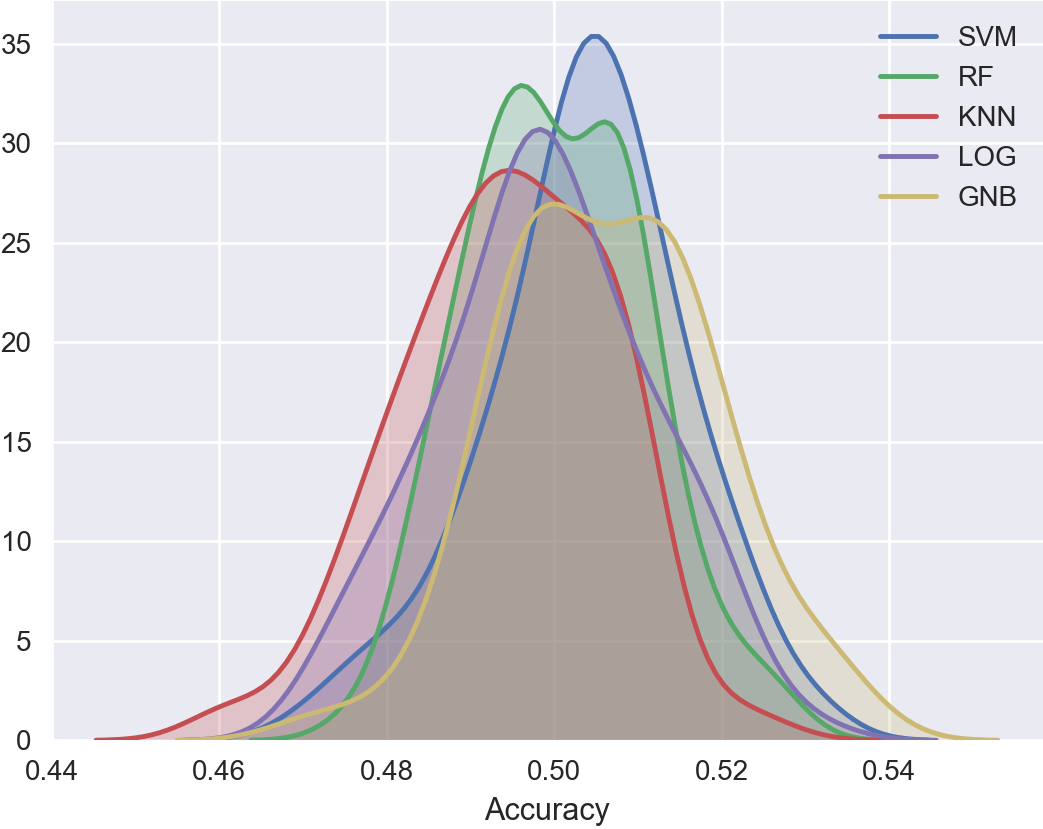}
\caption{Stock movement accuracy histogram of 100 randomly selected S\&P500 securities over 10 years by model.}
\end{figure}
\FloatBarrier

Logistic Regression is applied to test the same 100 randomly selected S\&P500 stocks utilized for econometric research of 20 years and tuned using optimal parameters found for all four stock indices 
. However, to verify general performance accuracy on these securities, all models generate predictions. Each model's parameters are also chosen by finding the best average Alpha across all indices. These results are stored, but ultimately only Logistic Regression predictions were used for trading 
.

Accuracy distributions for all stocks surprisingly find Logistic Regression to be among the least accurate models, while SVM and GNB performed the best. Performance differences between models were minute. It is difficult to ascertain if accuracy distributions would converge on a mean of 50\% if additional stocks were tested. We maintain it would. The variation in Figure 18 does not appear significant enough to have arisen from anything besides random chance
. 

Independent findings in Figure 18 diverge from the expected performance of Logistic Regression. Regardless of if accuracy converges about a normal distribution centered at 50\%, it underscores the possibility that Logistic Regression's performance with stock indices can not be replicated. Yet, perhaps Logistic Regression has the ability to consistently detect and trade appropriately in moments before high volatility, as was the case for index performance.

\FloatBarrier
\begin{figure}[ht]
\label{fig:RawPerf}
\centering
\includegraphics[width=1\linewidth]{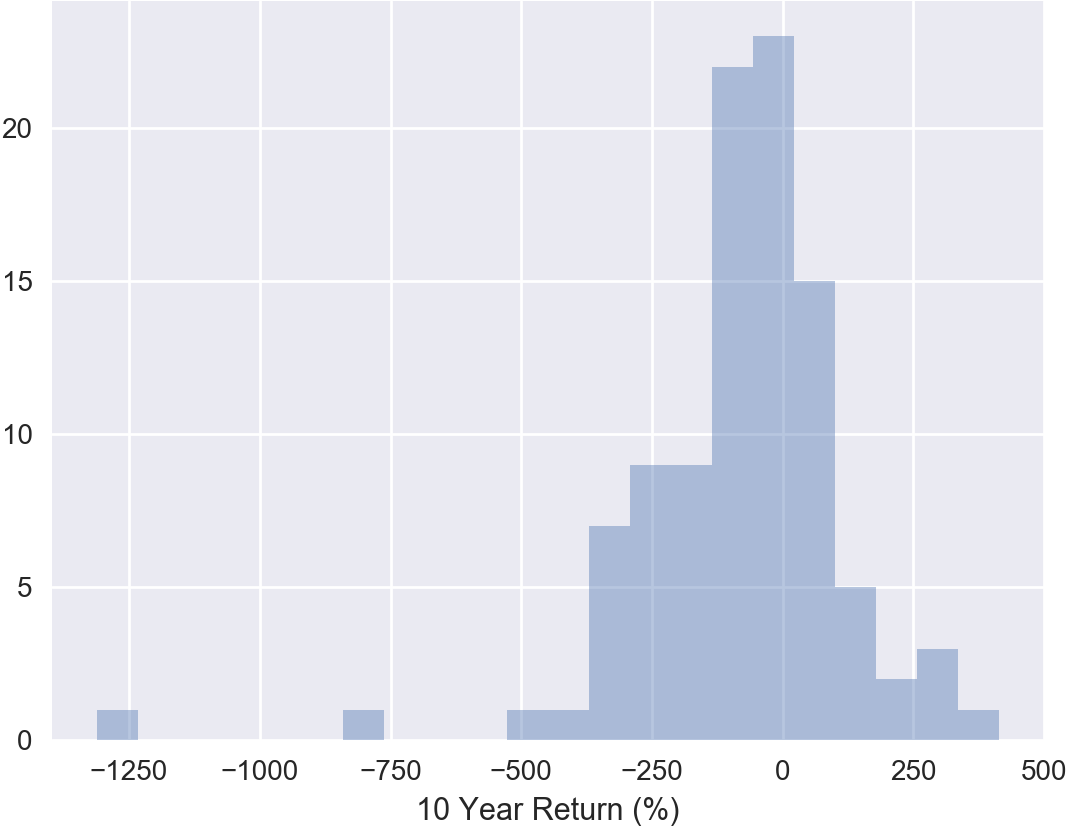}
\caption{Trading performance (Alpha) histogram of 100 randomly selected S\&P500 securities over 10 years for all models.}
\end{figure}
\FloatBarrier

Trading performance, outlined in Figure 19, refutes such a claim. Across all 100 securities, Alpha distributed around returns of 0\% with a slight left skew towards negative returns. The influence of transaction costs may generate left skew, but likely not to the extent seen in Figure 19. Returns depict strong evidence that Logistic Regression does not have a consistent means of predicting price movements or  achieving above-market returns.

Considered together, the experiments of this study starkly contrast the work of \cite{Patel2015}  and \cite{Dash2016} in Indian markets, \cite{Kara2011} in Istanbul, and \cite{Oliveira2013applying} in Brazil. With that said, all three of these markets host a small fraction of the trading volume and total market capitalization of United States markets. Therefore, we assert several hypotheses that may explain discrepancies between existing literature and our analysis. 
 
 First, perhaps the models utilized in this study were not sophisticated enough to garner above market returns. This potentiality is highly unlikely, as all implementation methods mirror those  of \cite{Patel2015}. One could argue that the time-frame analyzed was simply a more efficient time-period for markets, but this is ignored as it asserts market efficiency recently arose
 . Conversely, the United States financial markets are more efficient than most foreign markets. 
 
 The majority of machine learning studies that only utilize price data were conducted in developing or developed foreign markets much smaller than those of the United States. This factor could indeed be confounding findings, and should be further assessed by completing weak-form analysis in other large developed markets like China, Japan, or Europe
 . Third, previously published studies may have selectively disclosed data from their analysis such that it aligns with their claims. Infidelity is of particular concern for \cite{Patel2015}, who tests only four Indian securities with no independent verification. Nevertheless, the volume of studies with similar weak-form findings implies that results are at least partially valid for certain markets.
 
\subsection{Trend-specific Data Superior for Algorithm Training}

\FloatBarrier
\begin{figure}[!ht]
\label{fig:TrendNonTrendRet}
\centering
\includegraphics[width=1\linewidth]{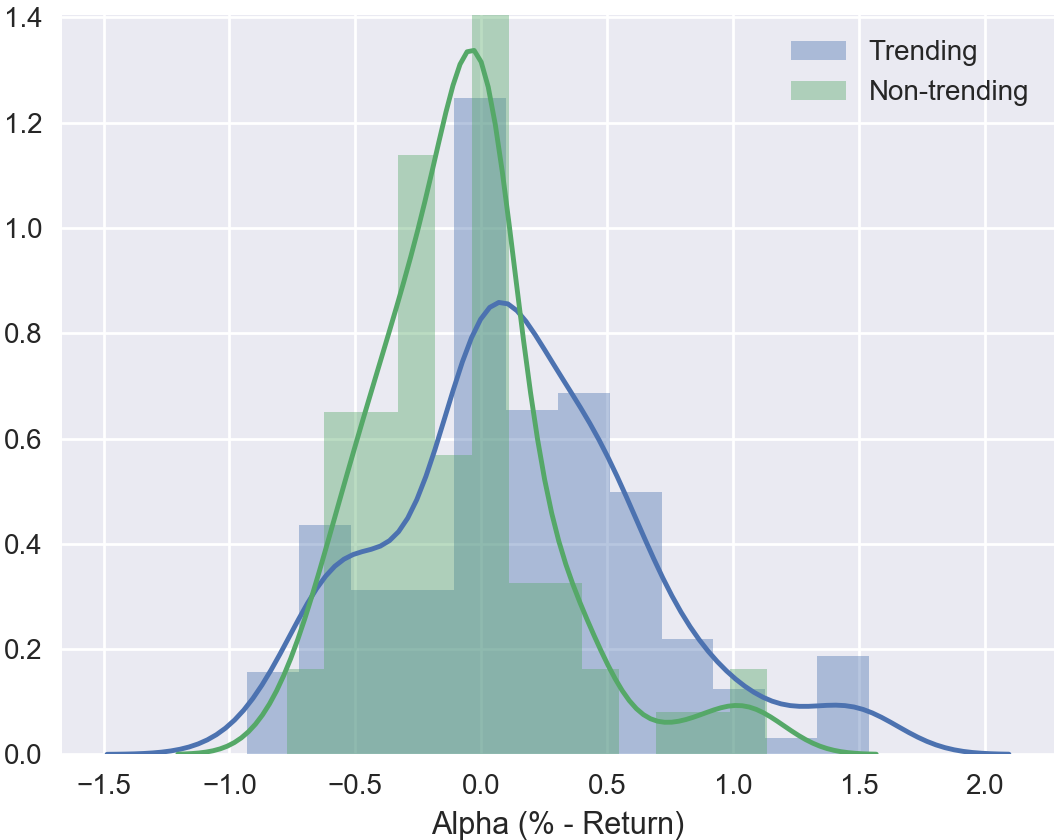}
\caption{Relative frequency histogram of trading returns (Alpha) for all trending and non-trending models. Alpha is given as percent return over 10 years.}
\end{figure}
\FloatBarrier

Trending and non-trending performance in this study mirrored the findings of \cite{Patel2015,Kara2011,Oliveira2013applying}. In Figures 11,13,15, and 17, non-trending data proved to be unable to properly train any models. Not a single algorithm consistently outperformed the market.
, and many did not trade more than once. As shown, the majority of algorithms appear to mirror the market exactly, taking and maintaining a buy-and-hold position. Closer inspection reveals that many models only outputted one trading decision over the entire ten years. Some, like GNB trading for RUT, spent nearly 10 years without trading at all. 

With continuous ratio data, the intuition behind the metric may be difficult for the computer to identify pro-grammatically. In fact, it is our assumption that the abnormal trading (or lack thereof) of models can be attributable to an inability to properly tune to a large, sparse feature space. Some algorithms like RF overcame this sparsity more easily, likely because of its stochastic, ensemble based prediction methods. However, more analytical models like Logistic Regression were rendered useless by continuous data.

Additionally, leaving technical ratios in their raw form increased model training and testing time substantially. In back-testing, Random Forest regressions took four times as long to complete with non-trending data. Non-trending Support Vector Machine tests were unable to be collected, as a single experiment took eight days to complete on our servers. Particularly in investing, where timing is essential to success for any active fund manager, algorithms must be able to operate efficiently. If there is any potential for above-market returns by leveraging technical data or otherwise, distilling ratios and indicators into the buy or sell message
they convey is the most effective method for preprocessing we have found in this study. Even so, we remain skeptical of methods that purport to generate above market returns by timing the market.


\section{Discussion}
\label{S:9}

The objective of this analysis was to jointly consider econometric and machine learning research as they pertain to weak-form Efficient Market theory. Economic and econometric research has transitioned from generally supporting of the validity of efficient markets to hotly debating its merits and claims. By contrast, machine learning research applied to predicting asset prices consistently finds success predicting market trends and garnering abnormal returns. Even today, when all EMH theories are debated in economics and computer science alike, the consistency with which publications achieve above-market returns is surprising.

Our study merges these fields of research in an effort to better understand the discrepancies in empirical findings. Econometric research is first conducted to determine if stock price movements appear random. Augmented Dickey-Fuller and Variance Ratio tests both seek to identify stationarity in data. Historically, ADF test results are consistently observed to support efficient market claims while VR tests more commonly refute them. Both of these tests are run on 100 randomly selected S\&P500 securities. ADF test results run counter to traditional findings, as every test statistic rested below a 1\% significance level. All hypothesis tests rejected the null in favor of stationarity. VR tests were more mixed, but still strongly implied stationarity with 70\% of p-values below a 1\% significance level.

Establishing the possibility for market trending and therefore predictive potential, the econometric findings were then supplemented by machine learning implementations. With a scalable, dynamic system, daily stock price moment predictions were created in a training-testing loop that refreshed daily. Predictions collected in this manner informed a trading system that made buy, sell, and hold decisions accordingly. Transaction costs as well as portfolio performance metrics were recorded. 

Four stock indices were tested using the aforementioned framework. A highly diversified asset, stock indices are a composite of hundreds (or thousands) of stocks, mitigating many of the risks associated with bankruptcy or company failure. Since the trading system implemented for experimentation only trades one security at a time, the inherent diversification of stock indices makes them particularly suitable for controlled experimentation. Though verified with several individual S\&P500 company stocks, empirical findings for stock indices are particularly insightful as they most directly align with the notion of generating above-market returns while holding risk constant. 

Econometric findings did not align with machine learning prediction in American markets. No algorithm was able to garner predictive accuracy significantly above 50\%, and while Logistic Regression garnered above-market returns for all stock indices
, independent verification with 100 randomly selected S\&P500 stocks was inconclusive. We maintain that the price-movement accuracy and trading performance indicate models trained on only price data do not have the predictive power necessary to generate abnormal returns.

With that said, our findings do not suggest that stock prices are random or represent all available information. The former claim was refuted by the results of our Augmented Dickey-Fuller and Variance Ratio tests, while the latter was not considered at all in this study. Random Walk theory and Efficient Market hypotheses are subtly distinct from one another, and findings for each must be carefully qualified. Regardless, the aggregate conclusions drawn from the results of our analysis compellingly affirm the validity of the weak-form Efficient Market Hypothesis.

Machine learning as a discipline is still evolving at what appears to be an accelerating rate. The capabilities of pattern-recognition software in the future are potentially limitless
. Therefore, research investigating both Random Walk and Efficient Market theories must continue. Moreover, this analysis exclusively examined weak-form efficient markets. The advent of sentiment analysis, natural language processing, and ancillary data that may inform price-movement occurrences should also be considered when testing semi-strong and strong form efficient markets. Fortunately, the insatiable urge for researchers and investors alike to increase the returns of their assets will further Efficient Market and Random Walk research in perpetuity. 

%
%
\clearpage
\bibliographystyle{acm}
\bibliography{EMHCitations.bib}

\end{document}